\documentclass[twocolumn]{aastex62}
\usepackage{amsmath}
\usepackage{hyperref}

\newcommand{\lx}{$L_{\mathrm{X}}$}
\newcommand{\lxo}{$L_{\mathrm{[OIII]}}$ --$L_{\mathrm{X}}$}
\newcommand{\lxorat}{$L_{\mathrm{X}}/L_{\mathrm{[OIII]}}$\ }

\newcommand{\dloiii}{$\Delta L_{\mathrm{[OIII]}}$}

\graphicspath{{./}{figures_pdf/}}

\shorttitle{AGNs Lacking Optical Emission Lines}
\shortauthors{Agostino et al.}

\begin{document}

\title{A New Physical Picture for AGNs Lacking Optical Emission Lines}
\author{Christopher J.\ Agostino}
\affil{Department of Astronomy, Indiana University, Bloomington, IN 47405, USA}

\author{Samir Salim} 
\affil{Department of Astronomy, Indiana University, Bloomington, IN 47405, USA}

\author{Sara Ellison} 
\affil{Department of Physics \& Astronomy, University of Victoria, Victoria, British Columbia, V8P 1A1, Canada}

\author{Robert Bickley} 
\affil{Department of Physics \& Astronomy, University of Victoria, Victoria, British Columbia, V8P 1A1, Canada}

\author{Sandra Faber} 
\affil{Department of Astronomy and Astrophysics, University of California, Santa Cruz, CA, 95064, USA}

 
\begin{abstract}

In this work, we use $\sim 500$ low-redshift ($z \sim 0.1$) X-ray AGNs observed by XMM-Newton and SDSS to investigate the prevalence and nature of AGNs that apparently lack optical emission lines (``optically dull AGNs''). Although 1/4 of spectra appear absorption-line dominated in visual assessment, line extraction with robust continuum subtraction from the MPA/JHU catalog reveals usable [OIII] measurements in 98\% of the sample, allowing us to study [OIII]-underluminous AGNs together with more typical AGNs in the context of the \lxo\ relation. We find that ``optically dull AGNs'' do not constitute a distinct population of AGNs. Instead, they are the [OIII]-underluminous tail of a single, unimodal \lxo\ relation that has substantial scatter (0.6 dex). We find the degree to which an AGN is underluminous in [OIII] correlates with the specific SFR or D$_{4000}$ index of the host, which are both linked to the molecular gas fraction. Thus the emerging physical picture for the large scatter seems to involve the gas content of the narrow-line region. We find no significant role for previously proposed scenarios for the presence of optically dull AGNs, such as host dilution or dust obscuration. Despite occasionally weak lines in SDSS spectra, $>80\%$ of X-ray AGNs are identified as such with the BPT diagram. $>90\%$ are classified as AGNs based only on [NII]/H$\alpha$, providing more complete AGN samples when [OIII] or H$\beta$ are weak. X-ray AGNs with LINER spectra obey essentially the same \lxo\ relation as Seyfert 2s, suggesting their line emission is produced by AGN activity.
 \end{abstract}


\section{Introduction} \label{sec:intro}


X-ray surveys at the turn of the century have revealed the existence of active galactic nuclei (AGNs) with quite powerful X-ray emission but without the optical emission line signatures associated with typical AGNs \citep{elvis1981, fiore2000, mushotzky2000, barger2001, comastri2002, brusa2003, szokoly2004, rigby2006, caccianiga2007, civano2007, cocchia2007, trump2009, trouille2010, tbt2011, trump2011a, trump2011b, koss2017}. These AGNs have very weak or no emission lines (i.e., only upper limits), including a weak or undetected [OIII]$\lambda$5007 line which is normally produced in the AGN narrow-line region (NLR). The general appearance of the spectra is that of early-type, absorption-line galaxies. Various terms have been used to describe these objects. In more recent literature, AGNs with weak or no emission lines are usually referred to either as `optically dull AGN' or as `X-ray Bright Optically Normal Galaxies' (XBONGs). In both cases `optical' refers to the emission lines, not the host galaxy continuum.

[OIII] luminosity is often assumed to be an isotropic indicator of AGN strength \citep{keel1994, kauffmann2003, brinchmann04}, and is used to estimate fundamental AGN properties such as the accretion rate. However, if line emission is for some reason diminished, despite luminous X-rays, the use of this popular indicator may be problematic, as pointed out by \citet{trouille2010}. If there is a substantial population of AGNs with no lines, the common optical selection techniques like the Baldwin-Phillips-Terlevich \citep[BPT,][]{bpt1981} diagram would have to be considered incomplete, and the studies that rely on clean samples, either of AGNs or non-AGNs, would be affected. For these reasons, and in order to advance the fundamental understanding of the physics of the AGN, it is important to establish the prevalence and nature of AGNs with underluminous emission lines.

The estimates of the prevalence of this putative population among type 2 AGNs vary widely in the literature: from no more than a few percent \citep{rigby2006}, to a range of 10-20\% \citep{trump2009}, and all the way to 60\% \citep{moran2002,caccianiga2007}. A related but more fundamental question is whether optically dull AGNs constitute a distinct population to begin with \citep{civano2007}, in the sense that there is some physically motivated dividing line with respect to ``normal'' AGNs or a quantity in which normal and optically dull AGN will produce a bimodal distribution. Going even further, it has been suggested that optically dull AGNs are just normal AGNs (i.e., AGNs with typical \lxorat ratios) that have upper limits on emission lines merely due to larger distances and the resulting observational limitations or biases \citep{yan2011}. 

A variety of potential scenarios have been put forward to explain the weakness or the absence of AGN-like line emission in optically dull AGNs: (a) lines are swamped (diluted) by the inclusion of a strong continuum light within the spectroscopic aperture \citep{moran2002}, (b) dust obscures the ionizing source with a large covering fraction so that the narrow-line region (NLR) does not get ionized \citep{barger2001, comastri2002, civano2007}, (c) dust in the host galaxy substantially attenuates the emission lines \citep{rigby2006}, (d) a radiatively inefficient accretion flow (RIAF) does not sufficiently heat the NLR to produce the lines \citep{yuan2004, hopkins2009, trump2009, trump2011a, trump2011b}, and (e) a complex structure present in the NLR results in a low covering factor which does not absorb enough ionizing photons \citep{trouille2010}. No consensus has yet been reached regarding which, if any, of these scenarios is dominant.

Many of the previous studies that focused on AGNs with weak lines were based on deep X-ray pointings in smaller fields, resulting in relatively distant samples ($0.3<z<1$) for which the ancillary spectroscopy of the depth needed to detect weak lines is challenging. Furthermore, because of the redshifting out of the optical range, many of these galaxies lacked observations of H$\alpha$ and [NII] that are needed for robust emission-line classification. Samples wherein many AGNs have only upper limits on line emission tend to suggest a dichotomy between ``normal'' and ``lineless'' AGNs, which may hide a more gradual distinction of intrinsic line strengths,m e.g., \citet{trump2009}. Although optically dull AGNs have been studied among the nearest AGNs as well, the small sample sizes pose challenges for the generalization of the results or for ascertaining whether these objects form a distinct population.

Our study aims to overcome some of these limitations by focusing on a relatively large and uniform dataset of low-redshift ($z<0.3$), X-ray selected AGNs. Such a sample is made possible by the combination of a very large catalog of XMM-Newton serendipitous observations cross-matched to SDSS spectroscopy---an approach previously adopted in \citet{hornschemeier2005}, \citet{caccianiga2007}, and \citet{pons2014}. Furthermore, low redshift samples provide the advantage of a rich array of ancillary data, including a catalog of physical properties of SDSS galaxies. These data facilitate a more extensive selection of AGNs (in particular, going below the $10^{42}$ erg s$^{-1}$ threshold for X-ray luminosity), and provide diagnostics needed for the investigations of the root causes of underluminous emission lines. Finally, at these lower redshifts luminous AGNs are expected to have NLR sizes (where [OIII] is emitted) that are comparable to the physical scales probed by the SDSS spectroscopic fibers, minimizing host contribution.

In order to determine how the optically dull AGNs fit into the bigger picture and test if there is a dichotomy with respect to the normal AGNs, it is useful to consider them in the context of the correlation between the [OIII] and X-ray luminosities, which is known to hold for normal AGNs \citep{heckman2005, netzer2006, panessa2006,bian2007,melendez2008, lamastra2009, georgantopoulos2010, lamassa2010, tanaka2012b, berney2015, ueda2015, azadimosdef2017, glikman2018, lambrides2020, esparza2020}. However, because they are significantly deficient in [OIII], optically dull AGNs are often left out of these samples. Previous work that has tried to study them through a prism of the \lxo\ relation or the \lxorat ratio were often faced with the challenges posed by large fractions of samples having only upper limits in [OIII] detection \citep{civano2007,trump2009,trouille2010,yan2011,smith2014}. In contrast, a relatively low redshift sample with SDSS spectroscopy significantly overcomes this limitation by providing usable measurements of [OIII] emission for nearly all X-ray selected AGNs.

The paper is organized as follows: in Section \ref{sec:data_methods}, we outline our sample selection and methods; in Section \ref{sec:results}, we present our main results including whether [OIII]-underluminous X-ray AGN constitute a distinct type of AGN and the implications of such a population on common optical emission-line diagnostics; in Section \ref{sec:dis_scat} we discuss what may be causing the deficiency in the line emission and in Section \ref{sec:conc} we summarize our findings.

\section{Data and Methods}  \label{sec:data_methods}
In this study, our main aim is to evaluate whether X-ray AGNs with weak or no optical emission lines form a distinct population of AGNs. Our data consist of X-ray detections and accurate star formation rates (SFRs), for AGN selection based on a non-stellar X-ray excess, and of optical emission lines, for the relationship between X-ray and [OIII] emission and for AGN classification.

\subsection{Data} \label{sec:data}
X-ray sources are extracted from the tenth data release of the fourth XMM-Newton serendipitous source catalog (4XMM-DR10, \citealt{webb2020}), which contains more than half a million unique sources identified in archival XMM-Newton observations. 4XMM-DR10 provides X-ray fluxes in 9 bands, spanning the energy interval 0.2-12 keV, computed assuming a power-law spectral model with a photon index $\Gamma = 1.7$. The XMM-Newton energy bands are somewhat mismatched with respect to the commonly used full X-ray band (0.5-10 keV) and hard X-ray band (2-10 keV), which we will use in this work. 
We therefore convert the fluxes in the 0.5-12 keV band (4XMM-DR10 Band 8 minus Band 1) to the 0.5-10 keV band by multiplying by 0.91, and 2-12 keV band (4XMM-DR10 Band 7) to 2-10 keV band by multiplying by 0.87. Both factors are based on $\Gamma=1.7$. We require that the S/N in the hard X-ray band is greater than 2, which effectively results in all sources having S/N$>2$ in the full band as well. We additionally consider the angular extent of the X-ray sources.

The measurements of optical emission lines come from the MPA/JHU catalog\footnote{\url{http://www.mpa-garching.mpg.de/SDSS/DR7/}} based on SDSS DR7 \citep{abazajian2009} and derived following \citet{tremonti2004}. The global galaxy properties---SFRs, stellar masses and stellar continuum dust attenuations come from the second release of the GALEX-SDSS-WISE Legacy Catalog (GSWLC-2 \footnote{\url{https://salims.pages.iu.edu/gswlc/}}, \citealt{salim2016, salim2018}) and were determined using UV/optical+IR SED fitting. The GSWLC contains three catalogs which are based on the depth of GALEX UV observations: Shallow (GSWLC-A2), Medium (GSWLC-M2), and Deep (GSWLC-D2), which cover 88\%, 49\%, and 7\% of SDSS, respectively. For the purposes of this work, we use the medium depth catalog, GSWLC-M2, because it provides more accurate estimates of host properties than the shallow GSWLC-A2, while providing a much larger sample than GSWLC-D2. Galaxies in GSWLC-M2, and therefore in our final sample, span a redshift range of $0.01<z<0.3$.

Optically dull AGNs have weak or undetected narrow optical emission lines, and are not found among type 1 AGNs. Therefore, we restrict our analysis to type 2 AGNs by removing the galaxies classified by the spectroscopic SDSS pipeline as `QSOs', which are effectively the sources with emission line FWHMs greater than 500 km s$^{-1}$ \citep{sdssdr12} i.e., a general population of type 1 AGN and not just the quasars.  

We matched sources in 4XMM-DR10 catalog to galaxies in GSWLC-M2 catalog with a 7'' search radius, following \citet{brusa2007} and \citet{lamassa2016}. For $\sim 2\%$ of cases where an X-ray source was within 7'' of multiple GSWLC-M2 sources, the GSWLC-M2 source with the highest $r$-band flux was adopted as a match. We found 712 X-ray sources matching GSWLC-M2. 

\subsection{X-ray AGN candidate Selection} \label{sec:xray_selection}

\begin{figure*}[t!]
            \epsscale{1.0}
            \hspace*{-0.5cm}\plotone{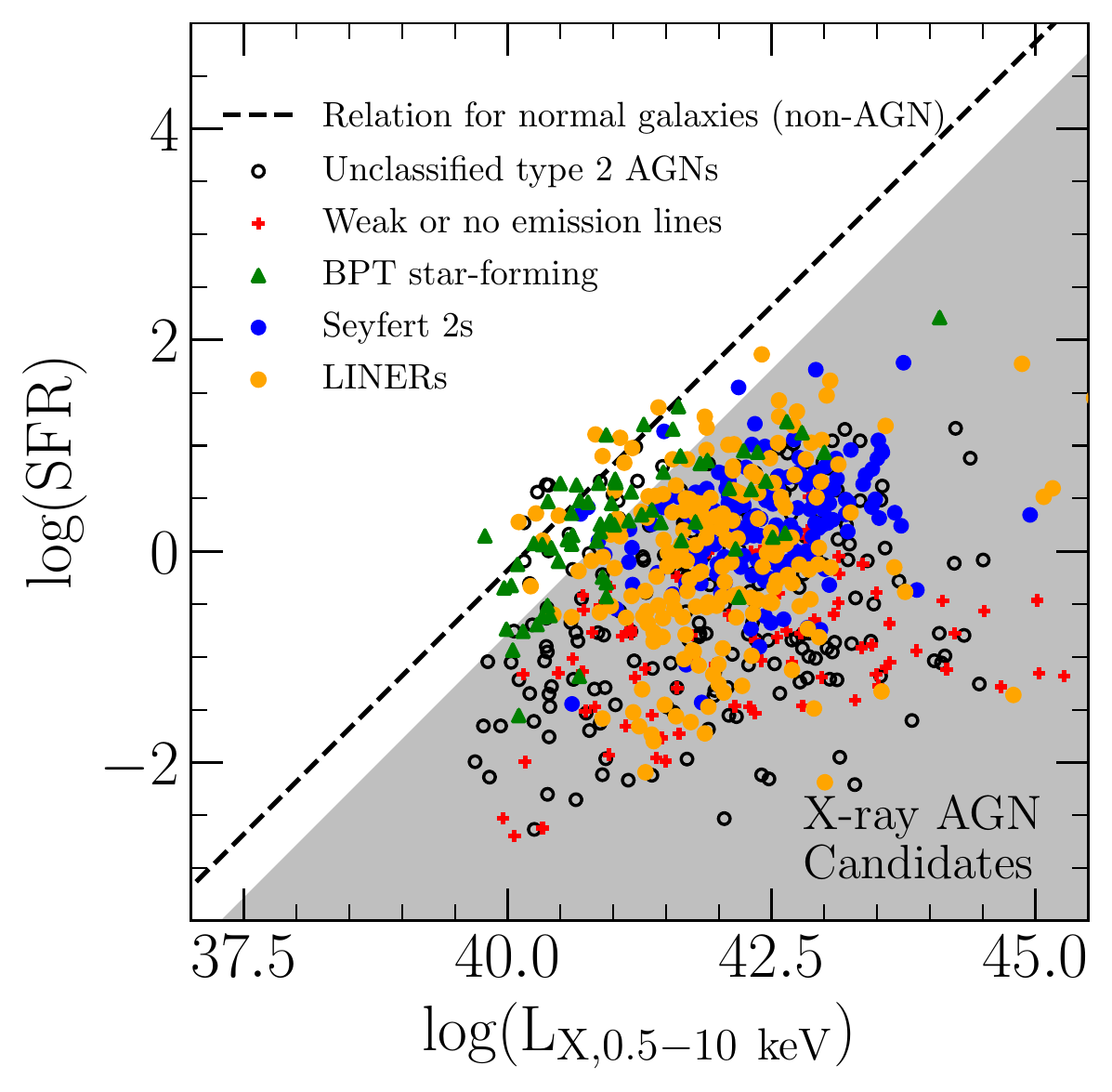}
            \caption{Selection of AGN candidates (shaded area) based on the X-ray excess. X-ray AGN candidates are objects that have an X-ray luminosity that is 0.6 dex greater than that predicted based on the SFR--\lx relation of non-AGN galaxies (black dashed line). Symbols correspond to the emission line classification of \citet{agostino2021}. LINERs and Seyfert 2s are shown as filled orange and blue circles, respectively. Unclassified type 2 AGNs (LINERs or Seyfert 2s) are shown as unfilled black circles. AGN candidates with weak emission lines or no detected lines are shown as red crosses. Galaxies classified as star forming (non-AGNs) according to the BPT are shown as green triangles.   \label{fig:lxsfr}}
        \end{figure*}

To select X-ray AGN candidates, we utilize the method described in \citet{agostino2019}, based on the correlation between SFRs and X-ray luminosity in nearby galaxies without AGNs (Figure \ref{fig:lxsfr}). We define AGN candidates as objects which have an excess greater than 0.6 dex in the X-ray luminosity compared to that predicted based on the SFR--$L_{\mathrm{X, 0.5-10\ keV}}$ relation. \citet{agostino2019} adapt the SFR--\lx relations of \citet{ranalli2003} provided separately for soft and hard bands into a single relation between the SFR and the full X-ray luminosity (in units of erg s$^{-1}$ here and throughout the paper unless noted otherwise), reproduced here:

\begin{equation}
L_{\mathrm{X, 0.5-10\ keV}} = \mathrm{SFR}/(0.66\times 10^{-40})
\end{equation}

The threshold of 0.6 dex corresponds to twice the standard deviation dispersion of the SFR--$L_{\mathrm{X, 0.5-10\ keV}}$ relation based on \citet{ranalli2003} data. The validity of this threshold is confirmed by observing that the galaxies in our sample lying above the relation in Figure \ref{fig:lxsfr} have a similar scatter. Of 712 X-ray sources matched to GSWLC-M2, 638 are selected as X-ray AGN candidates. We refer to them here as ``AGN candidates'' to allow for the possibility that in some galaxies X-rays originate from a source other than an AGN may dominate. 

Figure \ref{fig:lxsfr} shows that our X-ray AGN candidates span a wide range in X-ray luminosities ($40<\log L_{X}< 45$), with roughly 50\% below $\log L_{X}=42$, an often used threshold to select secure X-ray AGNs based on X-ray luminosity alone. Although selecting by $\log L_{X}>42$ is an excellent way to eliminate normal (non-AGN) if no information on SFR is available (see Figure \ref{fig:lxsfr}), it biases the sample against less powerful AGNs, reinforcing the notion that the majority of emission-line AGNs are Seyfert 2s. Historically, XBONGs/optically dull AGNs were defined as X-ray sources with $\log L_{X}>42$ \citep{georgantopoulos2005}, simply to ensure that they are not ``normal'' galaxies, but with the availability of SFRs, we can extend the investigation of the this phenomenon to low-luminosity AGNs.

 
Some studies that were published after \citet{ranalli2003}, e.g.  \citet{lehmer2010, lehmer2016, birchall2019, birchall2022},  have utilized stellar mass in addition to SFR in their estimation of the expected X-ray luminosity of non-AGN galaxies, on the basis that the stellar mass of a galaxy correlates with the emission from its low-mass X-ray binaries. AGN candidate selection using \citet{lehmer2010} or \citet{lehmer2016} relations results in similar samples as with our adopted method and our conclusions are not sensitive to it.

\subsection{Dust correction}

To carry out an investigation of the relationship between the intrinsic [OIII] luminosity and the X-ray luminosity, we perform dust corrections on the optical emission lines. 

When S/N in H$\beta$ is relatively high ($>$10), we use the Balmer decrement (assuming H$\alpha$/H$\beta$ = 3.1 and the \citet{cardelli1989} extinction curve) method to correct for the dust extinction. Otherwise, when H$\beta$ S/N is lower, resulting in a rather uncertain Balmer decrement, we follow \citet{agostino2021}, where the Balmer decrement is estimated from $A_V$ (stellar continuum dust attenuation). The estimation is based on the relationship between the Balmer decrement and $A_V$ for objects with well determined Balmer decrement (S/N in H$\beta>10$). In relatively rare cases where the Balmer decrement is formally greater than 3 or less than 0, we set it to be 3 and 0, respectively.

Ideally, one would also like to correct the X-ray fluxes for gas absorption, but to do so using the hardness ratios requires soft-band (0.5-2 keV) fluxes, which are usually not detected with sufficient S/N.\footnote{The lack of soft-band measurements does not affect the availability of full-band fluxes, because the latter are derived independently from the broadband measurements.} AGNs that are most subject to X-ray absorption tend to scatter above the \lxo\ relation, i.e., [OIII] luminosity is higher than expected (e.g., Figure 4 of \citealt{panessa2006}), and are not the focus of our analysis because optically dull AGNs have [OIII] luminosities much lower than what is expected based on their X-ray luminosity.

 \begin{figure}[t!]
            \epsscale{1.2}
            \hspace*{-0.5cm}\plotone{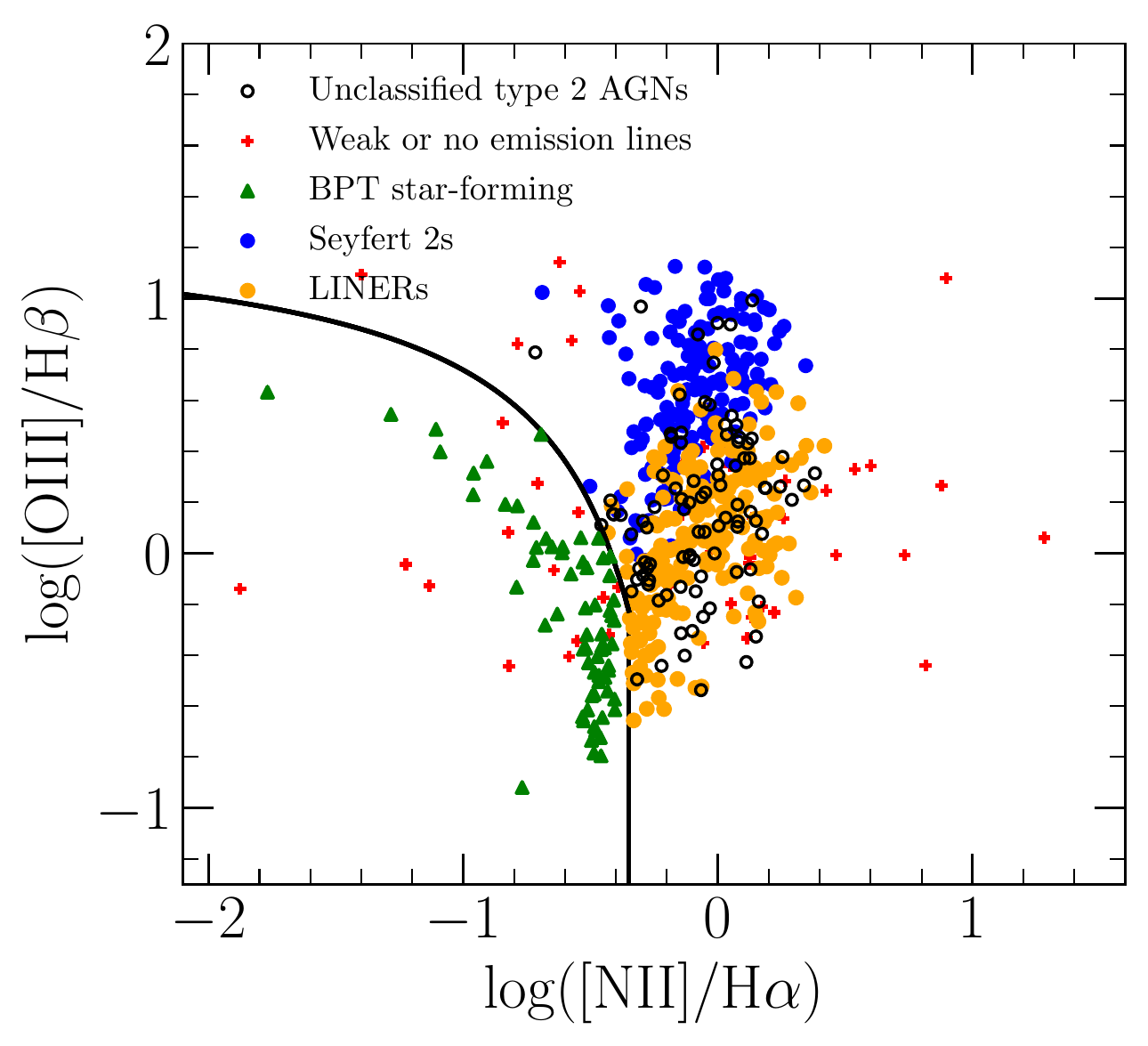}
            \caption{BPT diagram showing X-ray sources color-coded by the optical emission line classification scheme of \citet{agostino2021}. LINERs and Seyfert 2s are shown as filled orange and blue circles, respectively. Unclassified type 2 AGNs (LINERs or Seyfert 2s) are shown as unfilled black circles. AGN candidates with weak emission lines or no detected lines are shown as red crosses. BPT star-forming objects are shown as green triangles. The unclassifiable sources with weak or no emission lines are plotted without any S/N cuts. The solid black line is the modified \citet{kauffmann2003} line proposed by \citet{agostino2021} and which follows the \citet{kauffmann2003} line until log([NII]/H$\alpha$)=-0.35 where it becomes a 1-dimensional boundary beyond which objects are considered AGNs. \label{fig:bpt_subdiv}}
        \end{figure}

\subsection{Optical emission line classification}\label{sec:bpt_cl}

In this work, we consider whether X-ray AGNs with underluminous or apparently missing [OIII] emission form a distinct class, by studying them in the context of the \lxo\ relation. We will aid our sample selection and analysis with the optical emission line classification. To be classified using such diagnostic diagrams, one typically requires a minimum S/N in the emission lines used, and knowing whether or not an X-ray AGN candidate fails one or more of these S/N requirements provides useful information regarding the strength of their optical emission lines.

The most widely used scheme for selecting type 2 AGNs is the Baldwin-Phillips-Terlevich \citep[BPT, ][]{bpt1981} diagram, which uses [NII]$\lambda$6583, [OIII]$\lambda$5007, H$\beta$, and H$\alpha$. To identify AGNs based on optical emission lines, we follow the scheme provided by \citet{agostino2021}, and summarize it as follows. We require S/N$>2$ in all four  emission lines, and select as BPT AGNs those that lie above a modified version of the \citet{kauffmann2003} line. The modified demarcation line follows the \citet{kauffmann2003} line until log([NII]/H$\alpha)=-0.35$ at which point it becomes a 1-dimensional boundary above which objects are considered as AGNs regardless of [OIII]/H$\beta$ (hereafter [OIII] will always designate [OIII]$\lambda$5007). This modest modification provides a more complete AGN selection among the most massive galaxies, and is supported by the detailed analysis presented in \citet{agostino2021}. 

BPT-selected AGNs which have S/N$>2$ in [SII]$\lambda\lambda$6717,6731, [OII]$\lambda$3727, and [OI]$\lambda$6300 are further classified into Seyfert 2s and two types of LINERs (soft and hard) based on a clustering analysis of six or seven emission lines from \citet{agostino2021}. Furthermore, in cases where [NII] and H$\alpha$ have S/N$>2$ but either [OIII] or H$\beta$ has a low S/N and the usual BPT classification is impossible, we use the [NII]/H$\alpha$ ratio alone \citep{keel1985, stasinska2006} to classify as AGNs galaxies that have log([NII]/H$\alpha)>-0.35$. X-ray AGN which lack the requisite S/N even in [NII] and H$\alpha$ are considered `unclassifiable'.
Finally, galaxies that are classifiable by the BPT but lie below the demarcation line are referred to as BPT star-forming (SF). 

To summarize, all of our X-ray AGN candidates are subjected to optical emission line classification with one of the following outcomes (symbol used in Figure \ref{fig:lxsfr} is given in parentheses):
\begin{enumerate}
\item  Seyfert 2 (blue filled circles).
\vspace{-3mm}
\item LINER (orange filled circles).
\vspace{-3mm}
\item An AGN selected based on the BPT diagram or high [NII]/H$\alpha$ ratio for which Seyfert 2/LINER classification is not available (black open circles).
\vspace{-3mm}
\item Galaxy falling into the SF region of BPT diagram (green triangles).
\vspace{-3mm}
\item Unclassifiable (multiple weak/undetected emission lines) (red crosses). 
\end{enumerate}
We show X-ray sources (whether they have an X-ray excess indicative of an AGN or not) on a BPT diagram in Figure \ref{fig:bpt_subdiv}. We see that unclassified AGN are typically found in the region dominated by the LINERs. 

We see in Figure \ref{fig:lxsfr} that Seyfert 2s are more often found in galaxies with higher SFRs and higher X-ray luminosities, whereas LINERs span a wider range both in SFRs and X-ray luminosities.  BPT-SF galaxies are primarily near the SFR--\lx relation of non-AGN galaxies and represent many of the sources within 0.6 dex of the relation, consistent with the expectation that most do not contain an AGN and further justifying the use of that particular SFR--\lx relation and threshold for X-ray excess. Unclassifiable AGNs (weak or no lines) are found mostly at lower SFRs, as expected. 

X-ray AGNs that are ``misclassified'' as SF by BPT exist in our sample (green triangles in the shaded region of Figure \ref{fig:lxsfr}), but are uncommon ($\sim5\%$ of all X-ray AGN candidates). Some studies refer to them as `elusive AGN' \citep{smith2014,pons2014}. \citet{agostino2019} deals with ``misclassified'' AGNs in detail, and provides evidence that in the absence of SF they would most likely belong to the unclassifiable category, i.e., they are not normal AGNs subjected to some sort of SF dilution, but rather their intrinsic AGN lines would not be strong enough for a BPT classification in the first place. Thus, misclassified AGNs are not really incorrectly classified---they are essentially optically dull AGNs with central star formation. They will not be included in the remainder of the analysis because their emission lines are dominated by star formation. This gives the final size of the X-ray AGN candidate sample of 600.  

When we say that optically dull AGNs lack emission lines, what exactly do we mean? Which line or lines are considered? What is meant by a line not being present? Is it a low flux (luminosity), or low S/N (detectability)? Often the optically dull AGN assignment is based on the visual appearance of a spectrum having mostly absorption lines. In this paper we avoid such uncertainties by analyzing our sample of X-ray AGN candidates in terms of [OIII] measurements, without assigning the XBONG/optically dull label to any particular subset.

\subsection{Relationship between [OIII] and X-ray luminosities}

In order to identify X-ray AGNs that are potentially underluminous in [OIII], we must adopt a reference \lxo\ relation that will tell us what [OIII] luminosity is expected for an AGN with some X-ray luminosity . 
In this work, we adopt the \lxo\ relationship from \citet{panessa2006}:
    \begin{equation}\label{eq:panessa}
    \log L_{\mathrm{X, 2-10\ keV}} = \log L_{\mathrm{[OIII]}}\cdot 1.22 -7.55
\end{equation}
    which has a dispersion of 0.59 dex (computed from the $L_{\mathrm{X, 2-10\ keV}}$ and  $L_{\mathrm{[OIII]}}$ values in Table 2 of \citealt{panessa2006}). The relation was derived using 45 nearby (median distance $\sim$26 Mpc) optically selected Seyferts 1 and 2 from \citet{ho1997} (referred to as the `Tot' sample in Table 3 of \citealt{panessa2006}). Note that by very virtue of selecting standard Seyferts their sample did not include optically dull AGNs. A very similar slope (1.23) and scatter (0.61 dex) was found in the analysis of \citet{berney2015} using a larger sample (321) of Seyferts 1 and 2.

\section{Results} \label{sec:results}

In this section, we consider the optical emission-line properties of X-ray AGN candidates. In Section \ref{sec:underlum}, we focus on the [OIII] properties of the X-ray AGN candidates in order to determine if there is a distinct population of [OIII]-underluminous X-ray AGN. In Section \ref{sec:bpt_res}, we discuss the completeness of optical emission-line diagnostic classifications with respect to the X-ray selection.

\subsection{Fraction of [OIII]-underluminous X-ray AGN} \label{sec:underlum}


AGNs with weak or absent line emission may indicate collectively the existence of a distinct type of AGN---the purported XBONGs/optically dull AGNs. Such a population would be challenging to study using the \lxo\ relation because optically dull AGNs will, by definition, have no [OIII] measurements or only upper limits, especially for deep-field samples at larger distances (e.g., \citealt{trump2009,yan2011}). However, a particular strength of a study based on SDSS spectroscopy of relatively local AGN ($z\sim 0.1$) is that almost all of X-ray AGN candidates have useful [OIII] measurements.

\begin{figure}[t!]
        \begin{center}
            \epsscale{1.2}
            \hspace*{-0.2cm}\plotone{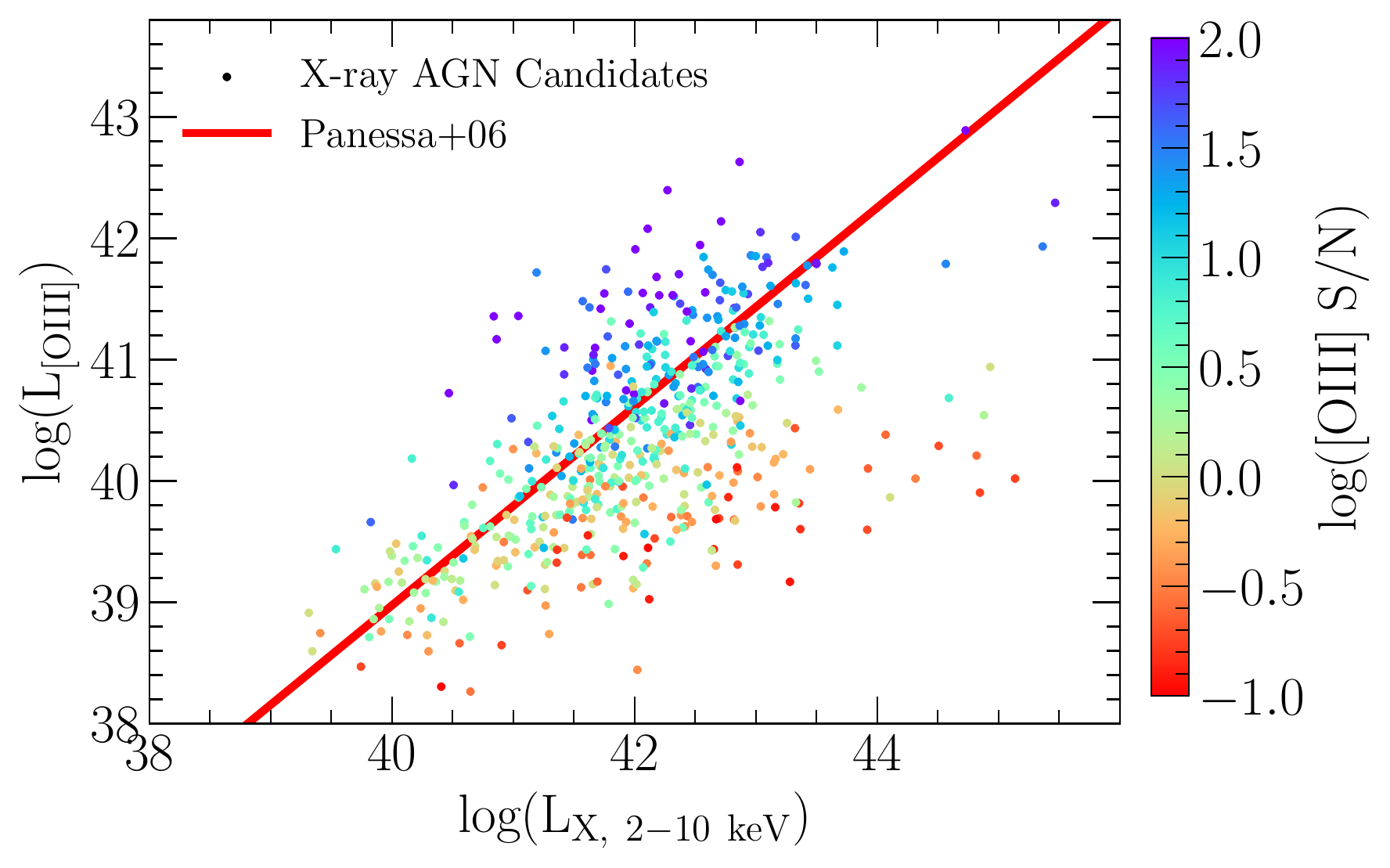}
            \caption{[OIII] luminosity versus X-ray luminosity with points color-coded by the logarithm of their [OIII] S/N. Only X-ray AGN candidates with [OIII] S/N$>1$ are shown. The median [OIII] S/N is $\sim13$. There is a substantial scatter around the \citealt{panessa2006}) relation (red line) established based on well-studied nearby Seyferts. Only 6\% of the sample is not shown because their [OIII] is entirely unconstrained (S/N$<1$). 
            \label{fig:lxo3}}

        \end{center}
        \end{figure}

In the Figure \ref{fig:lxo3}, we show the relationship between [OIII] and X-ray luminosities of all X-ray AGN candidates. We color code each point by their [OIII] flux S/N. The \citet{panessa2006} relation matches well the general trend of our sample. Importantly, only 37 out of 600 objects cannot be included in the plot because their [OIII] is entirely unconstrained (S/N$<1$). Sources with [OIII] S/N$<1$ span X-ray luminosities from $40.4<logL_{X}<44.5$, similar to the rest of the sample. We confirm that the placement of objects with nominally low SNR ($1<$S/N$<3$) is not random (Figure \ref{fig:lxo3}), as would be the case if their [OIII] emission was indeed unconstrained, but rather they are preferentially [OIII]-underluminous. 

We can indeed see that some X-ray AGN candidates lie considerably below the \lxo\ relation---more than 1 dex. However, for such sources (and consequently optically dull AGNs) to form a distinct population of AGN and not just be the tail of a ``normal'' distribution, one could imagine that they would be concentrated below the \citet{panessa2006} relation and/or that there would be a large fraction of sources omitted from the plot because they have [OIII] S/N$<1$. Although visually it does not appear that there are multiple populations present, there is a larger number of sources below the \lxo\ relation than above it, and so we turn to more formally determine whether or not there is a distinct [OIII]-underluminous population of AGN.

        \begin{figure}[t!]
        \begin{center}
            \epsscale{1.2}
            \hspace*{-0.2cm}\plotone{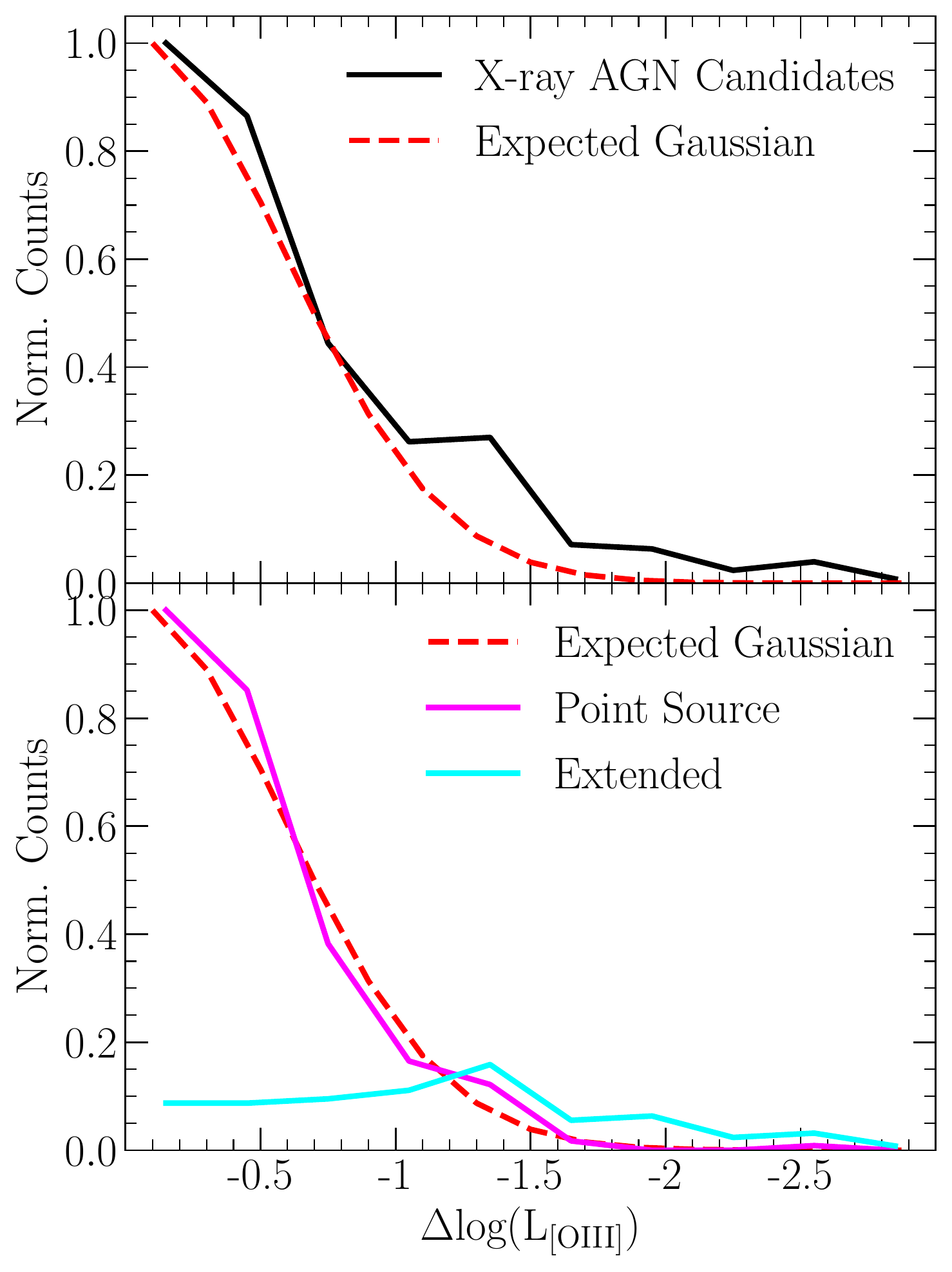}
            \caption{Offsets in [OIII] luminosity from the \citet{panessa2006} \lxo\ relation of the X-ray AGN candidates lying below the relation (below red line in Figure \ref{fig:lxo3}). Top panel shows all X-ray AGN candidates (black solid line). The expected Gaussian with a dispersion of 0.59 dex found for nearby AGNs in \citet{panessa2006} is shown in both panels as a red dashed line. The point sources and extended (resolved) X-ray sources are shown separately in the bottom panel. The ``excess'' population in the upper panel is the result of a contamination from the resolved sources, where X-rays arise primarily from halo hot gas rather an AGNs. \label{fig:deltalo3}}

        \end{center}
        \end{figure}

The fraction of [OIII]-underluminous objects (i.e., optically dull AGNs) among all X-ray AGN candidates can be estimated by studying the distribution of offsets from the \lxo\ relation ($\Delta L_{\mathrm{[OIII]}}$). The excess in the number distribution of offsets over the normal distribution would indicate the existence of a separate population. We refer to the abundance of such population as the {\it underluminous fraction} and designate it $f_{\mathrm{O3UL}}$.

We determine the underluminous fraction by counting the number of X-ray AGN candidates that lie below the \lxo\ relation ($\Delta L_{\mathrm{[OIII]}}<0$) and comparing it to what would be expected from a Gaussian with the same peak as the actual distribution and with a standard deviation of 0.59 dex \citep{panessa2006}. This is shown in the top panel of Figure \ref{fig:deltalo3}. We see that the distribution of X-ray AGN candidates follows the Gaussian distribution up to $\sim$ 1 dex below the \lxo\ relation, but diverges considerably further below. Using this simple approach, we compute the underluminous fraction as

\begin{equation}
    f_{\mathrm{O3UL}} =\frac{ND+O-G}{T}
\end{equation}

where $G$ is the number in the Gaussian component (the expected number if all objects belong to a single population), $ND$ is the number of non-detections (flux in [OIII] has S/N$<1$), $O$ is the actual number of objects, and $T$ is the total number of X-ray AGN candidates. Both $G$ and $O$ refer to objects with $\Delta L_{\mathrm{[OIII]}}<0$.
With this, we find $f_{\mathrm{O3UL}}=0.17\pm0.03$. We have estimated the error on this fraction by varying the number of counts in each bin (with width of 0.3 dex) using Poisson sampling wherein the average value is the number of counts in that bin and the standard deviation the square root of that number. This Poisson sampling is done for each bin separately, and they are summed up to determine $O+ND$, where $ND$ is treated as a bin of its own and Poisson re-sampled in the same way. We repeat this exercise 1,000,000 times to obtain the error estimate on the $f_{\mathrm{O3UL}}$.

        \begin{figure}[t!]
        \begin{center}
            \epsscale{1.2}
            \hspace*{-0.2cm}\plotone{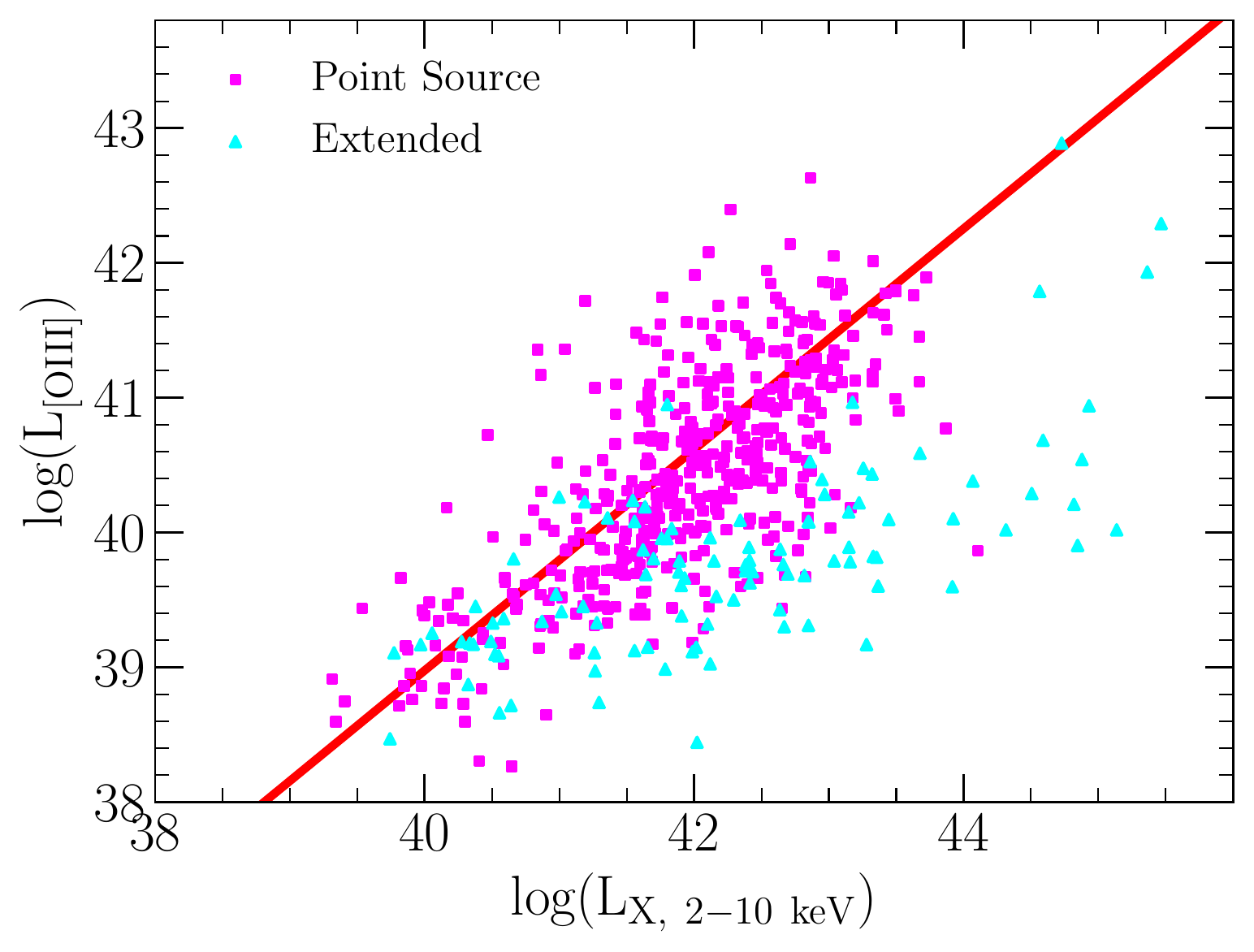}
            \caption{[OIII] luminosity versus X-ray luminosity. The X-ray AGN candidates are split into point sources (magenta squares) and extended ones (cyan triangles). The extended sources are primarily found among the [OIII]-underluminous population, so they can mimic XBONGs/optically dull AGNs if the extent of the X-ray source is not known.  \citealt{panessa2006}) relation (red line) is based on well-studied nearby Seyferts. Most of the higher X-ray luminosity (log$L_{X}>43$) sources are extended. A more thorough investigation into the differences of extended sources and point sources is provided in the Appendix.
            \label{fig:lxo3_ext_noext}}

        \end{center}
        \end{figure}

Based on the above analysis one would conclude that 15-20\% of all X-ray AGNs are [OIII]-underluminous and therefore suggestive of the existence of a separate population. However, an important factor we have not considered so far is the potential that in some objects the X-ray emission is fundamentally not associated with an AGN, but instead is a result of hot gas emission from the halos of massive galaxies, groups, and clusters. Such X-ray emission will be extended, unlike point-source emission of AGNs. In the 4XMM-DR10 catalog, sources are considered extended (more accurately, resolved) if the size of the source is greater than 6'' (PSF size of XMM-Newton). The extent of the sources are reported in 4XMM-DR10 either as 0'' for unresolved sources or the measured size above 6'' for resolved ones.

In our sample of relatively low-redshift sources the distinction between point sources and extended emission can be easily achieved with XMM-Newton resolution, but this may be more challenging at higher redshifts. In Appendix \ref{app:ext} we outline some alternative strategies to identify contaminating objects.

Of our 600 X-ray AGN candidates, 126 must be removed because they have a non-zero extent in 4XMM-DR10. In Figure \ref{fig:lxo3_ext_noext}, we show the \lxo\ relation with our sample color-coded by whether or not they are extended X-ray sources. A substantial fraction of the objects which lie well below the \lxo\ relation are extended, and most of the highest luminosity ($\log(L_{X})>43$) X-ray sources are extended. Their luminosities are typical of the X-ray luminsotities associated with clusters  \citep[$43<\log(L_{X})<46$;][]{mulchaey2000}. 
We now repeat the analysis of the distribution of offsets from the \lxo relation. Without extended sources, the distribution of X-ray AGNs is actually consistent with being normally distributed, as shown in the bottom panel of Figure \ref{fig:deltalo3}. With extended sources removed, we find the [OIII]-underluminous fraction to be f$_{\mathrm{O3UL}}$=0.025$\pm$0.03. In other words, there is no evidence of a \textit{distinct} population of AGN lacking emission lines. Rather, the AGNs with weak lines or no detectable lines lie in the tails of the (quite wide) distribution of [OIII] luminosities that is inherently present at every X-ray luminosity. This is to first order equivalent to saying that \lxorat distribution is very broad, as pointed out by \citet{trouille2010}.

\subsection{Intrinsic dispersion of \lxo\ relation}\label{sec:int}
With the removal of the extended sources, the Gaussian with a dispersion of 0.59 dex (dispersion found in \citealt{panessa2006}) apparently describes the distribution of [OIII]-underluminous X-ray AGNs well. Direct computation of the RMS of sources below the \lxo\ relation gives a consistent answer of 0.60 dex.

The median error in the hard X-ray luminosity is 0.1 dex and it is 0.03 dex for the [OIII] luminosity, suggesting that the observed large scatter of 0.6 dex is intrinsic and not inflated by observing errors.

\subsection{Implications for optical emission-line selection of AGNs}\label{sec:bpt_res}

\citet{agostino2019} reported that $\sim$60\% of what they considered to be X-ray AGNs at $z<0.3$ had significant signal (with SNR$>$2) in all 4 BPT lines to allow selection using that method, indicating a relatively high level of incompleteness of BPT selection of AGN. However, that work did not consider the X-ray extent of the sources and it did not impose a S/N cut on the X-ray fluxes, allowing spurious sources to be included. A more nuanced treatment of the completeness of AGN selection using SDSS optical emission lines is presented here.

\begin{figure}[t!]
        \begin{center}
            \epsscale{1.22}
            \hspace*{-0.3cm}\plotone{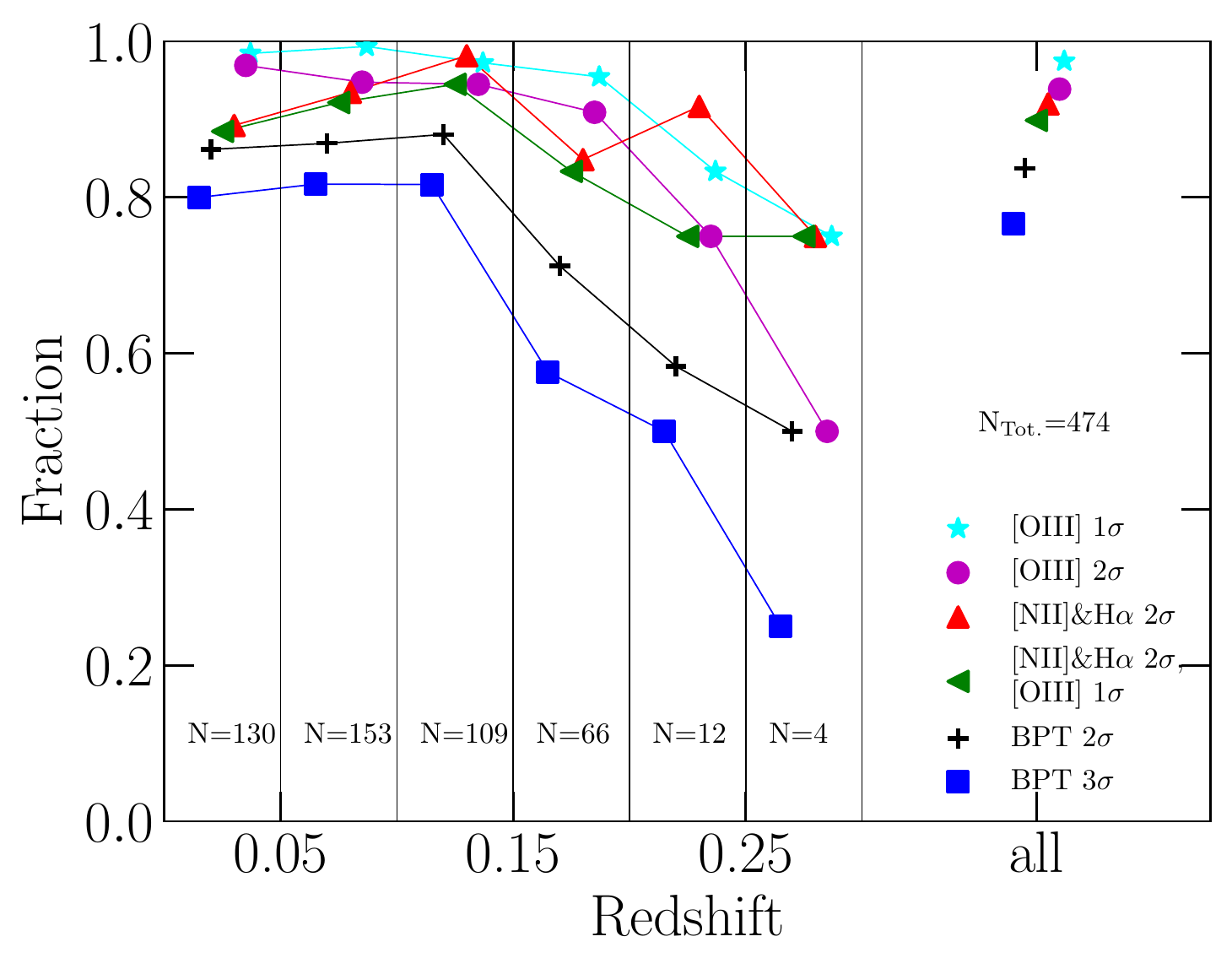}
            \caption{Fractions of X-ray AGNs that pass different emission line S/N requirements. The overall fractions (regardless of the distance) are shown in the rightmost column. Different lines represent different line thresholds---from the weakest requirement that only [OIII] is detected at $1 \sigma$ level to all four BPT lines being detected at $3 \sigma$. As expected, the detectability drops significantly with the distance, in particular above $z=0.15$. BPT identifies 80-90\% of X-ray AGNs at $z<0.15$. \label{fig:fclass}}

        \end{center}
        \end{figure}

We now focus on 473 of the 600 X-ray AGN candidates that are unresolved in X-ray photometry and which we consider to be genuine AGNs based on X-ray excess (Fig. \ref{fig:lxsfr}). In Figure \ref{fig:fclass}, we show the detectability (pertaining to SDSS) of different combinations of emission lines as a function of distance. The overall fractions (regardless of the distance) are shown in the rightmost column. Different lines represent different line thresholds---from the weakest requirement that only [OIII] is detected at $1 \sigma$ level to all four BPT lines being detected at $3 \sigma$. As expected, Figure \ref{fig:fclass} shows that the detectability drops significantly with the distance, in particular above $z=0.15$. Regardless of the distance, 78\% of X-ray AGN can be selected using the SDSS BPT diagram requiring SNR$>$3 in all 4 lines. This fraction is few percent higher (83\%) if requiring a lower threshold of SNR$=2$, as we did in \citet{agostino2019}. Dropping the requirement for [OIII] detection to just 1 sigma, but requiring the [NII] and H$\alpha$ to be detected (so that their ratio indicates an AGN) raises the detectability to 90\%. Overall, we conclude that the AGN selection using SDSS lines is rather complete at $z<0.15$: it is around 85\% using BPT, and 95\% when BPT selection is appended by using the high [NII]/H$\alpha$ ratio as an indication of AGN when either H$\beta$ or [OIII] have SNR$<2$. 

Figure \ref{fig:fclass} also shows the fraction of X-ray AGN with any [OIII] emission to be 97\% overall (as expected from analysis in Sec \ref{sec:underlum}). In closer bins that fraction approaches 100\%, again highlighting that fundamentally there are no AGNs lacking line emission altogether. At z$>$0.2 the fraction drops to 80\%, i.e., when AGNs seemingly have no [OIII], it is due to the sensitivity of observations.



   \section{Discussion} \label{sec:discussion}

\subsection{AGNs with weak emission lines as an integral part of the regular AGN population} \label{sec:dis_main_res}

In this study, we have investigated the emission-line properties of [OIII]-underluminous X-ray AGN. These objects are equivalent to XBONGs, or `optically dull' AGNs, as referred to in previous studies, except that we are able to identify them among the less luminous (log \lx $<42$) AGNs as well. We find the term `[OIII]-underluminous' preferable to these other names for several reasons. First, the term `optical' on its own evokes emission dominated by the continuum rather than optical emission lines. Second, by referring to luminosity and the quality of being underluminous, we move from observed qualities (whether a line is detected or not, which depends on a specific instrumental setup) to intrinsic physical properties of the entire ensemble of AGNs. 

It is perhaps for these and some other conceptual reasons that the answer to the question ``what fraction of AGNs are XBONGs or optically dull?" has eluded a clear answer. Assuming no extended X-ray sources masquerading as AGN are present (a potential problem at higher redshifts), the issue is further complicated by the fact that some studies consider the XBONG/optically dull category to encompass all X-ray AGNs that do not look like strong Seyfert 2s, that is, they include AGNs with HII-like spectra and LINERs. Such accounting can bring the fraction of AGN with ``atypical" spectra as high as 60\% \citep{moran2002,caccianiga2007,goulding2009}, mostly because of the inclusion of LINERs (we discuss the issue of LINERs in Section \ref{sec:dis_liners}).

\begin{figure}[t!]
        \begin{center}
            \epsscale{1.2}
            \hspace*{-0.2cm}
            \plotone{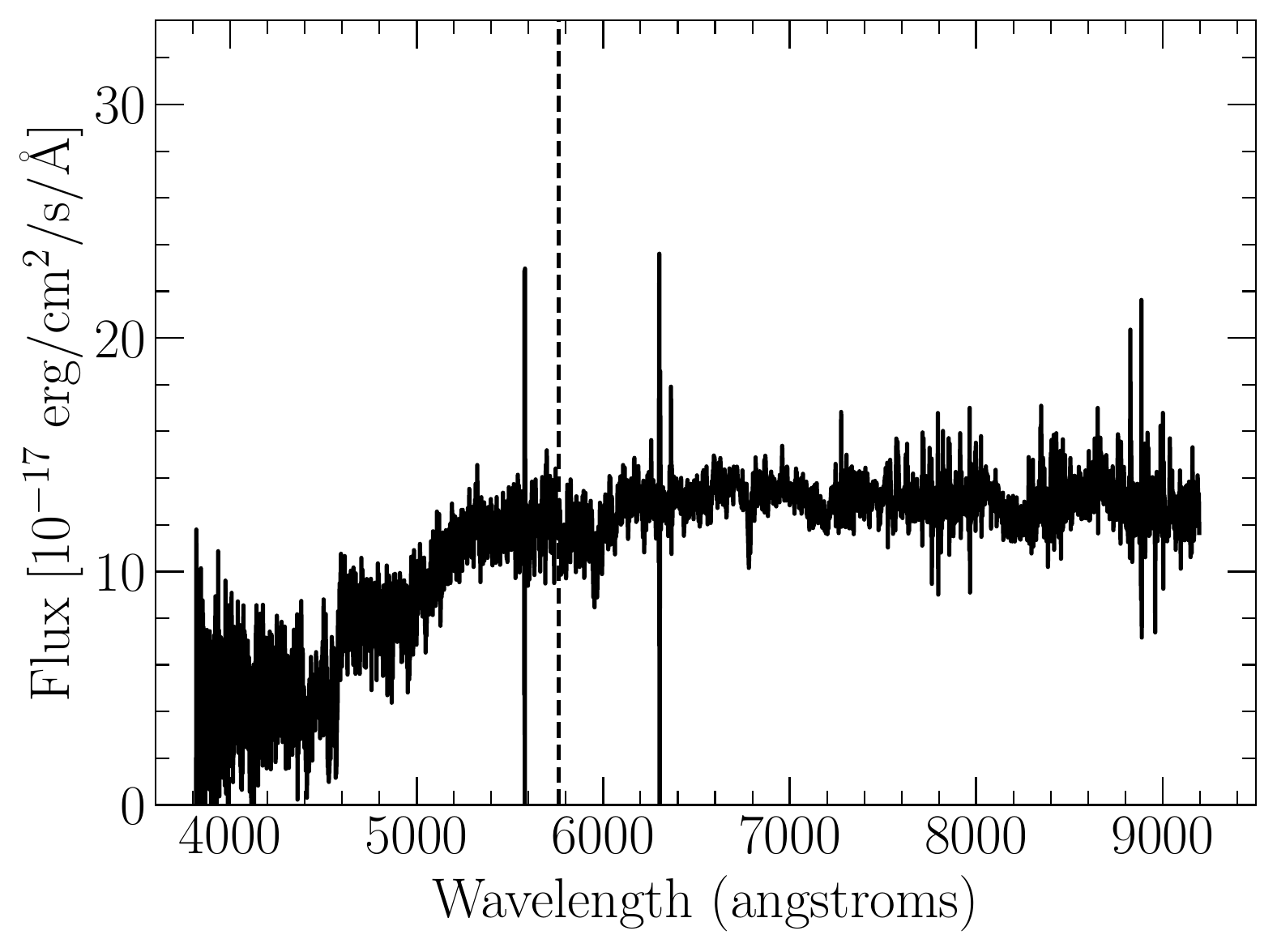}
            \plotone{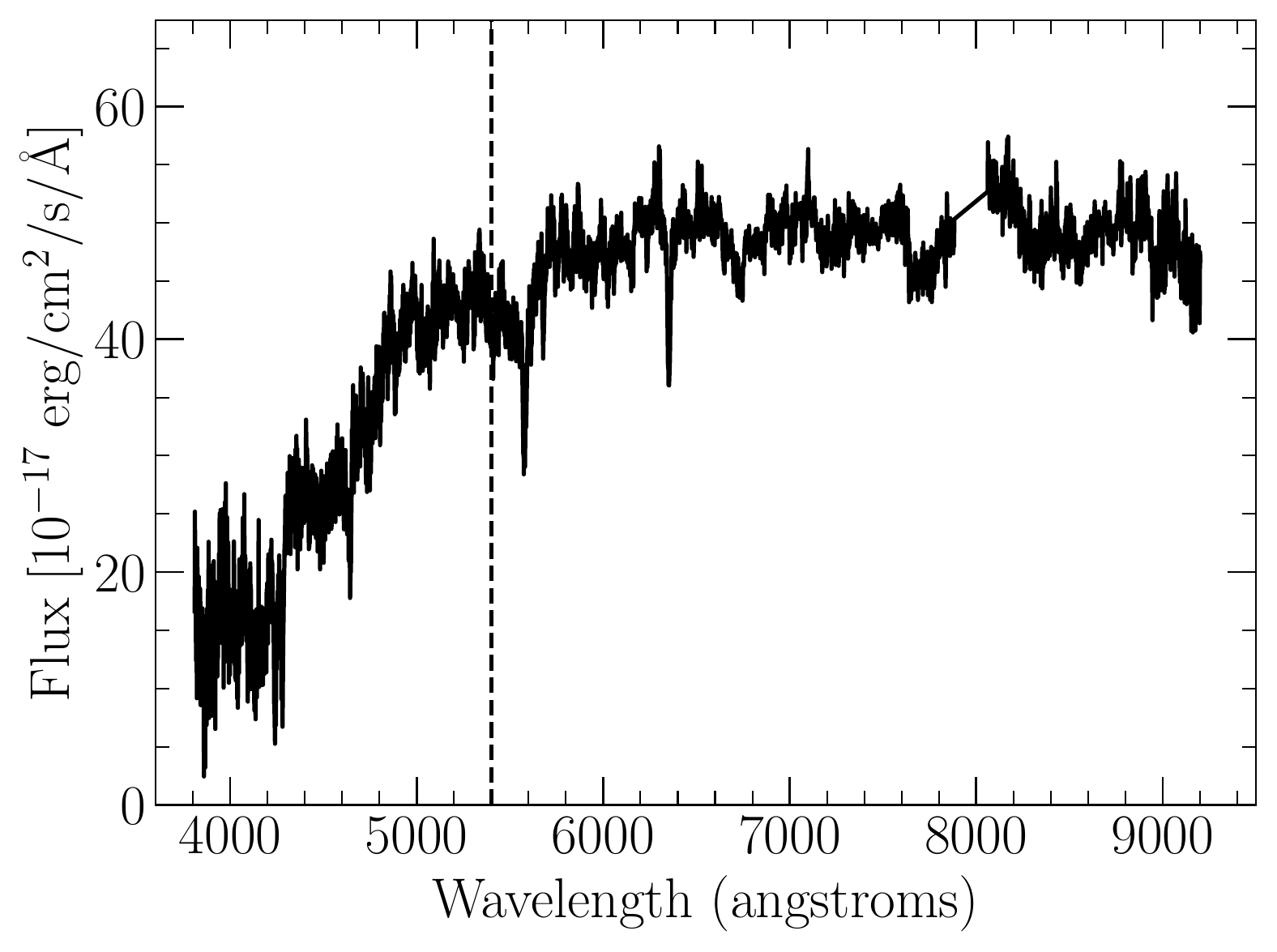}
            
            \caption{Two SDSS spectra of AGNs with high X-ray luminosities ($>10^{42}$ erg/s) and low [OIII] S/N ($\sim3$). The first source has a redshift of 0.1508 and the second 0.0778, and the [OIII] lines are at $\sim5760$ \AA\ and $\sim5400$ \AA, respectively, and their locations are shown with a dashed line.
           These spectra demonstrate that even a well-measured emission line (S/N$\sim$3) will not be obvious in the observed spectra and can only be revealed through the detailed removal of host continuum.
            \label{fig:o3_sn3_spec}}

        \end{center}
        \end{figure}

When restricted to X-ray AGNs with no obvious lines (absorption-line spectra), the estimates of the XBONG fraction range from a few percent \citep{malizia2012,rigby2006}, to around 20\% \citep{barger2002,georgantopoulos2005}. \citet{trump2009} quote a range 10-20\%. However, without a strict definition of what a spectrum with ``no lines'' means, the question is moot. To illustrate this point, only 2\% of X-ray AGN in our sample have [OIII] measurements that are so low to be entirely unconstrained. On the other hand, when we subject the spectra in our sample to a visual inspection, 24\% \textit{appear} not to have [OIII] emission. To make this point salient, we present two example spectra of X-ray AGNs with log(L$_{X}$)$>$42 and with [OIII] S/N$\sim$3 in Figure \ref{fig:o3_sn3_spec}. Emission lines (especially [OIII]) are not obviously detected in visual assessment and their presence is only revealed when the host continuum is removed. As discussed in Section \ref{sec:bpt_cl}, a categorical assignment to a class of ``AGN with no emission lines'' is conceptually problematic, and it is more useful to consider continuous physical quantities, such as the [OIII] luminosity, and to avoid conclusions based on the visual assessment of observed spectra. 

More fundamentally, the question ``what fraction of AGNs are XBONGs or optically dull?'' should be asked only if there is a separate population of such AGN to begin with. Whether XBONGs are a distinct class has been identified as ``the main issue'' by \citet{civano2007}. We find that no such distinct population exists. Rather, AGNs are intrinsically characterized with a very broad distribution of line luminosities even at fixed X-ray luminosity.  

If XBONGs or optically dull AGN are just the tails in the \lxorat distribution, why has the idea that they form a distinct population been so persistent? It may be that the notion of what a typical AGN should look like is skewed by the fact that we examine the spectra visually on \textit{linear} flux scales. An AGN that falls only 1.5 $\sigma$ below the \lxo\ relation will have lines 10 times weaker than the AGN on the \lxo\ relation, and one is naturally under an impression that such a spectrum is ``atypical'' for an AGN, but rather continuum/absorption-line dominated. 

\subsection{Sources of scatter in the \lxo\ relation} \label{sec:dis_scat}

While there may not exist a distinct population of AGNs that are underluminous in emission lines, there is nonetheless a wide range of [OIII] luminosities at a given X-ray luminosity. Therefore, we rephrase the often posed question ``what is the nature of optically dull AGNs?'' to a more fundamental question ``why is the scatter in \lxo\ relation so large?''. In doing so, we are following the spirit of the approach of \citet{trouille2010}, who pointed out the very large range in the \lxorat ratio of AGNs. We prefer to conceptualize the question in terms of the \lxo\ relation, rather than the \lxorat ratio so as to not have to assume that a relation is strictly linear.

First, let us emphasize that we are primarily concerned with the scatter below the \lxo\ relation ([OIII] less luminous than implied by the relation). There is a considerable amount of scatter above the \lxo\ relation as well, especially for $\log L_{\mathrm X}>41.5$, but its cause is well understood: most of it arises from the gas absorption of X-ray emission close to the AGN \citep{panessa2006}, in some cases in the Compton-thick regime. It is important to point out that we see a diminished flux of X-rays because our sightline happens to pass through very high column of gas. However, the NLR ``sees'' the ionizing source through typically non-obscured sightlines. Therefore it is only the X-ray luminosity that needs to be corrected for the absorption, whereas the [OIII] luminosity is unaffected. The reduction in scatter in the \lxo\ relation achieved by the application of a correction to X-ray luminosities can be appreciated by comparing the two panels of Figure 4 in \citet{panessa2006} (note that their \lxo\ plots have flipped axes with respect to ours.) However, even with the correction, the scatter is relatively large (0.59 dex, the value that also describes distribution of our [OIII]-underluminous AGNs; Figure \ref{fig:deltalo3} and Section \ref{sec:int}).

Focusing now on the scatter below the \lxo\ relation we note that previous works have explored a number of factors which might affect the optical line emission in the anomalous AGNs. The key to revealing if any of those factors is actually relevant in producing underluminous AGN is to compare [OIII]-underluminous AGN with the ones with more typical [OIII] luminosities. Thus, we revisit these investigations by looking at a number of parameters as a function of the distance from the \lxo\ relation: $\Delta L_{\mathrm{[OIII]}}$. 

We also note that one potential source of dispersion in the \lxo\ relation would be the variability of the X-ray flux. However, this effect is expected to be quite small. Although the majority of AGNs show some variability in X-ray flux, the average amplitude is only 15\% \citep{mateos2007} , or 0.06 dex, much smaller than the intrinsic dispersion. 

\subsubsection{Aperture effects}\label{sec:dis_aperture}

Beyond the physical processes that may drive the scatter, it is crucial to remember the observational limitations brought about by the SDSS fibers having an aperture size of 3'', which corresponds to $\sim5$ kpc at $z=0.1$. \citet{moran2002} presented the case that aperture effects may be behind the apparently atypical (non Seyfert 2-like) spectra of AGN in deep X-ray fields. They observed how the nuclear spectra of nearby Seyfert 2s change when encompassed to cover the entire galaxy (integrated spectra). It is important to point out that the aperture effects manifest in two flavors: (1) if the inclusion of host light contains the emission from star-forming (HII) regions, they may ``dilute" otherwise normal AGN lines (star formation dilution), and (2) the inclusion of host \textit{continuum} light \citep{comastri2002} may overwhelm otherwise normal AGN lines (continuum swamping), without adding lines from HII regions. The first phenomenon, star formation dilution, could result in the change of emission line ratios such that an AGN becomes classified as a star-forming galaxy in the BPT diagram. Indeed, that is one of the explanations for the ``misclassified'' AGNs also known as ``elusive'' AGNs \citep{maiolino2003,goulding2009,pons2014,agostino2019}. Interestingly, in \citet{moran2002}, only 2 out of 18 galaxies undergo this change when the classification is based on integrated spectra as opposed to nuclear ones. Recently, \citep{agostino2021} has shown that the shifts in the BPT diagram induced by the inclusion of HII regions in SDSS fibers are relatively small, so SF dilution of otherwise normal-strength AGN lines is not likely to be a major source of uncertainty of classification using SDSS emission-line diagnostic diagrams.

\begin{figure}[t!]
            \epsscale{1.2}
            \plotone{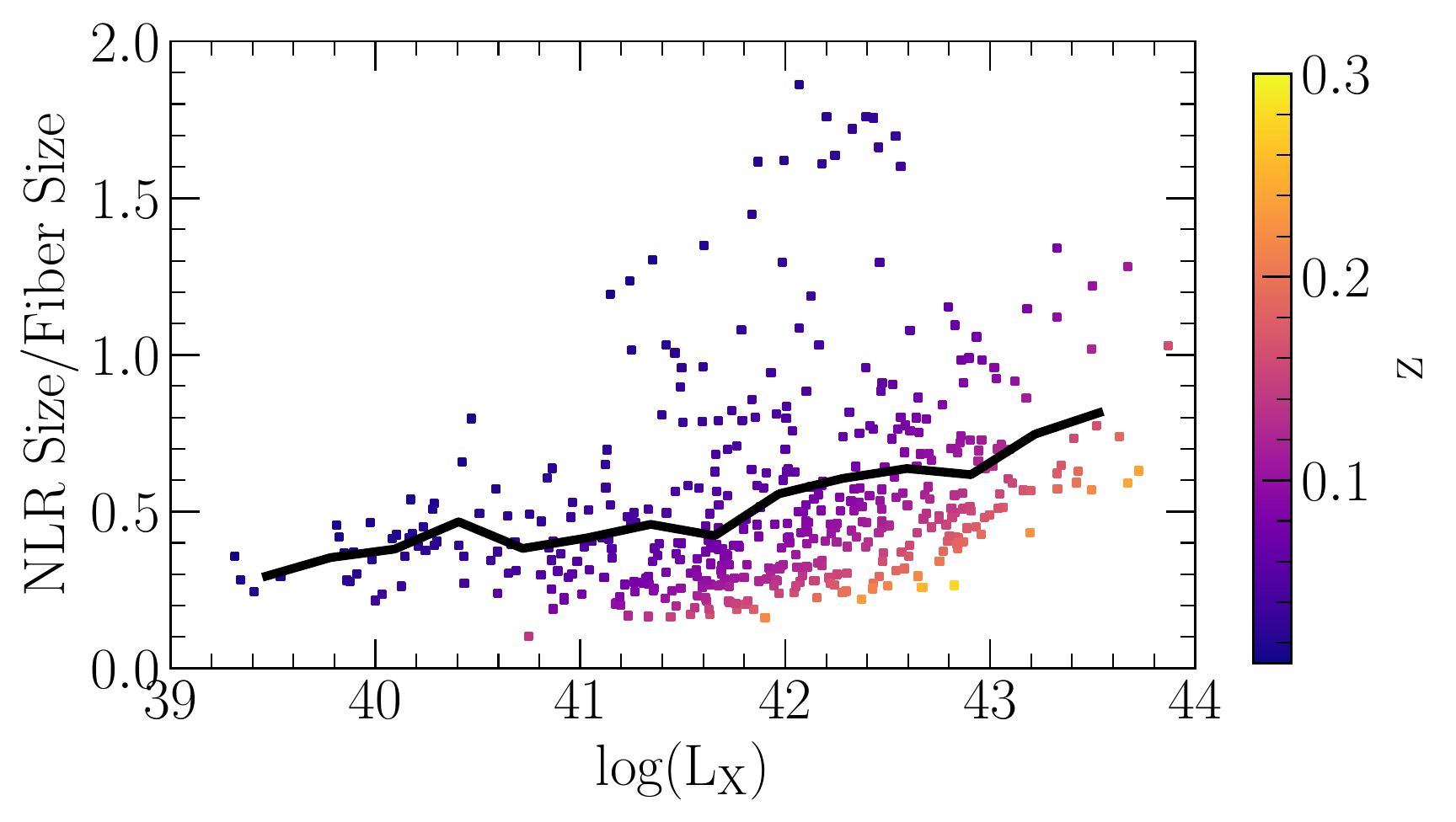}%
            \caption{Ratio of the size of the narrow-line region to that of the spectroscopic fiber versus the X-ray luminosity. Points are color-coded by redshift. A line showing the average trend is shown in black. NLRs occupy a significant portion of the fibers, which explains the modest degree of apertue effects.
            \label{fig:nlr_sizes}}

        \end{figure}

The root cause of the modest effect of SF dilution may lie in the size of NLR regions \citep{bennert2006a_nlr, bennert2006b_nlr, greene2011_nlr, liu2013_nlr, liu2014_nlr, law2018_nlr, chen2019nlr} actually being comparable in size to SDSS fiber (ranging from $300 pc$ to $5 kpc$ for the range of [OIII] luminosities in our sample), even at $z\sim 0.1$, which means that it would be more difficult to perturb the real NLR emission than that of a putative point-source. To directly test this hypothesis, we have estimated the NLR radius for each of our X-ray AGNs. To do so, we convert the measured X-ray luminosity into the expected [OIII] luminosity using the \citet{panessa2006} relation and then use the empirically determined relationship between [OIII] luminosity and NLR radius from \citet{chen2019nlr}. We show the resultant ratio of the NLR size to fiber size in Figure \ref{fig:nlr_sizes}. At log$L_{X}=42$, the average ratio of the NLR size to the extent covered by the fiber is approximately 0.5 for AGNs in SDSS. This plot showcases that the size of the NLR is not so minuscule, and in the absence of other sources of optical emission lines that may genuinely be stronger than the AGN, ought to dominate in SDSS spectroscopy.

       \begin{figure*}[t!]
            \epsscale{1.2}
            \includegraphics[height={2.3in}]{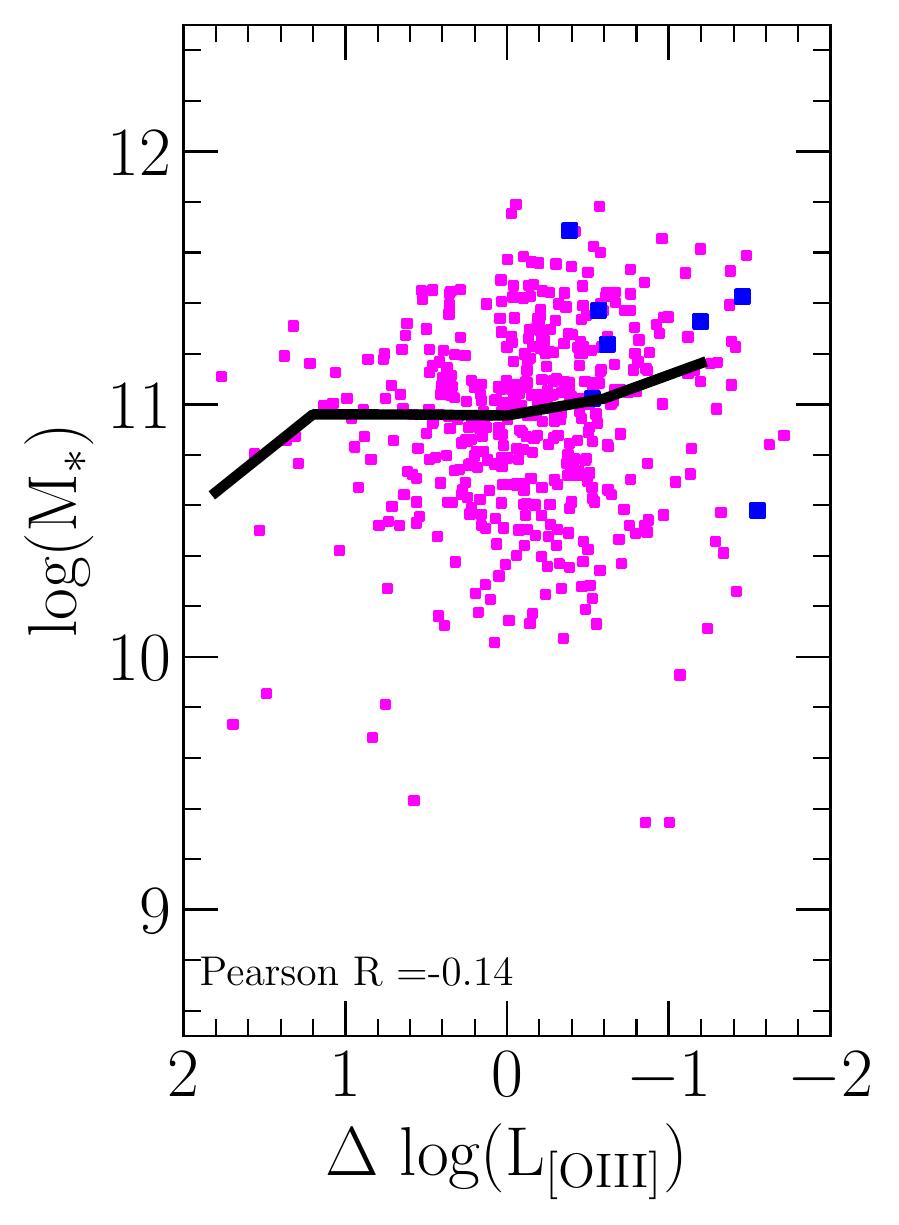}%
            \hfil 
            \includegraphics[height={2.3in}]{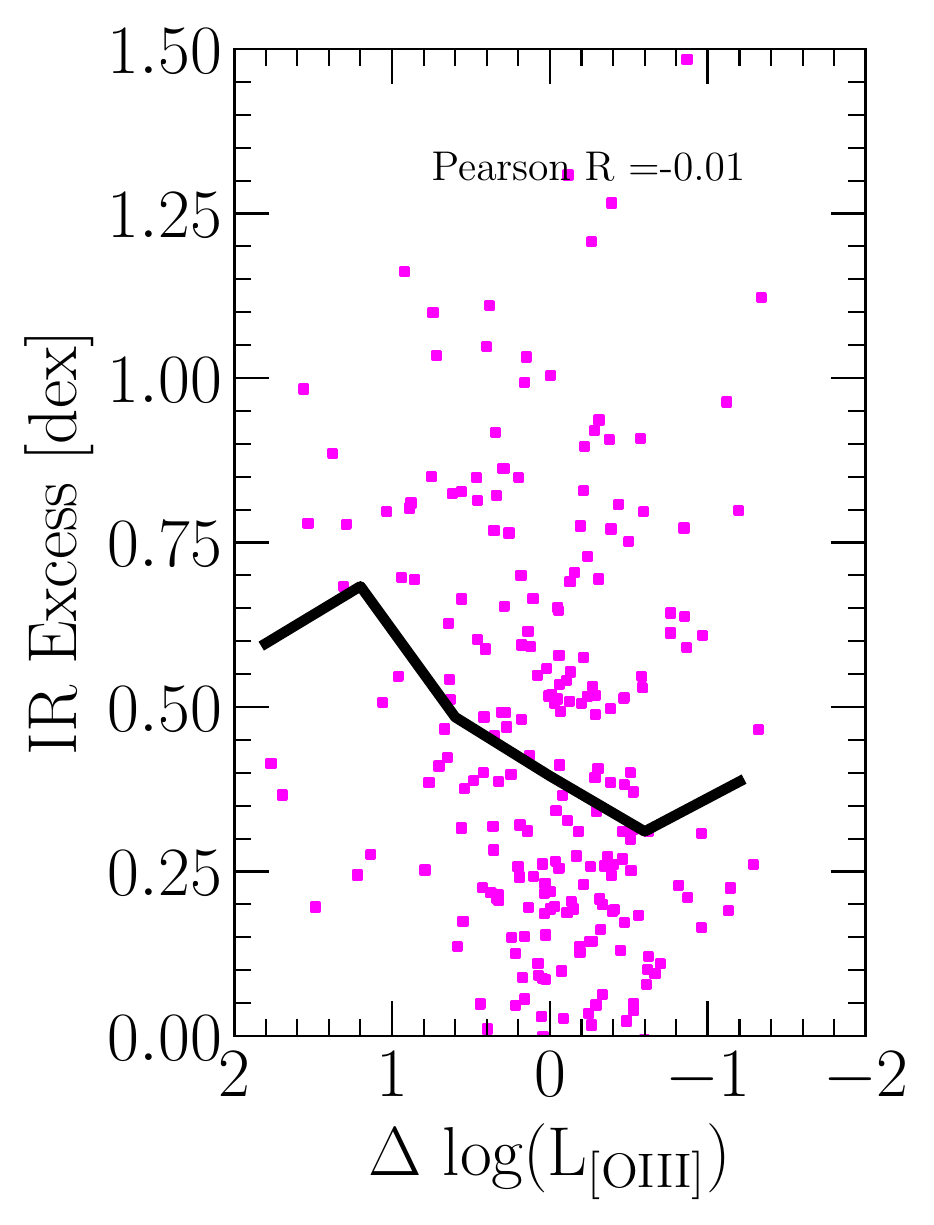}%
            \hfil
            \includegraphics[height={2.3in}]{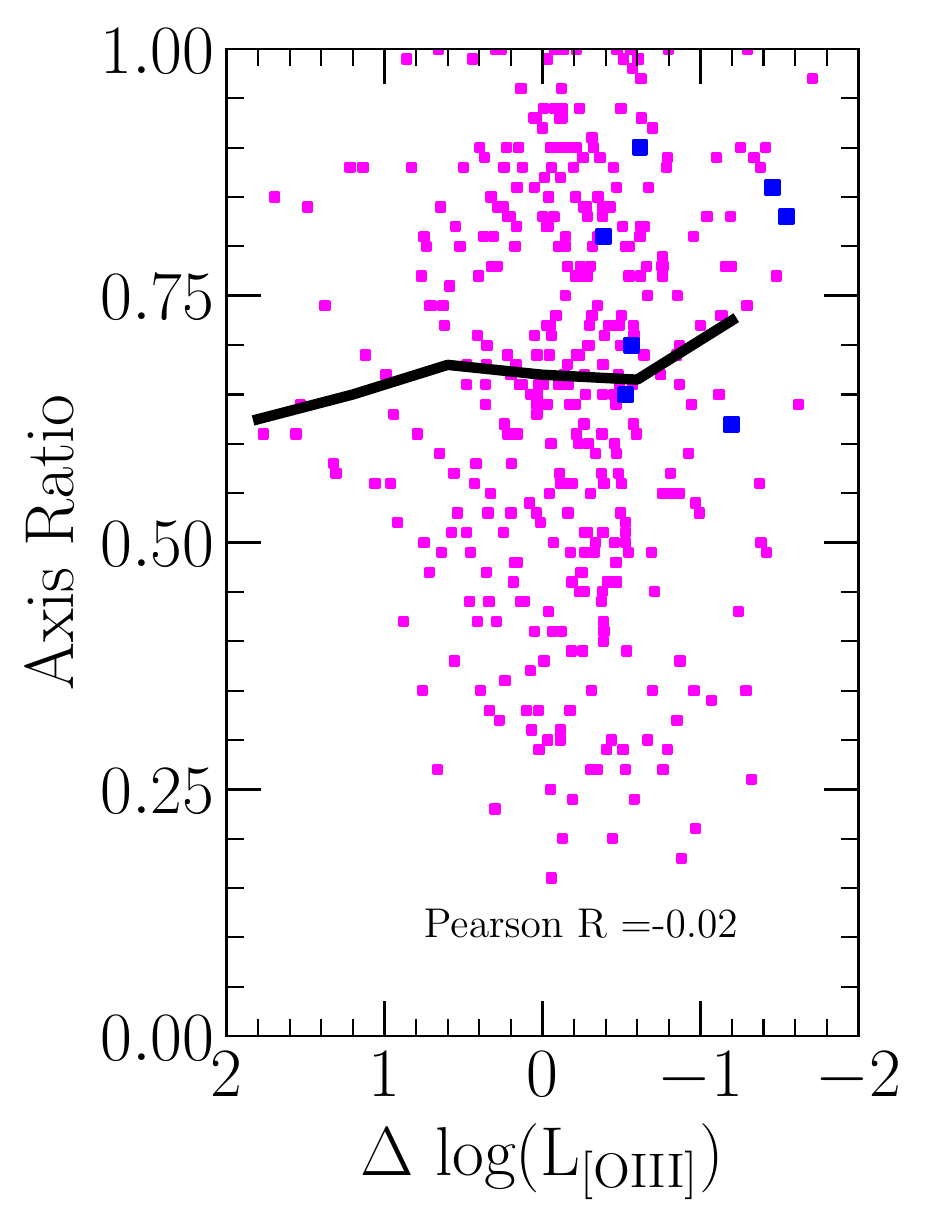}%
            \hfil 
            \includegraphics[height={2.3in}]{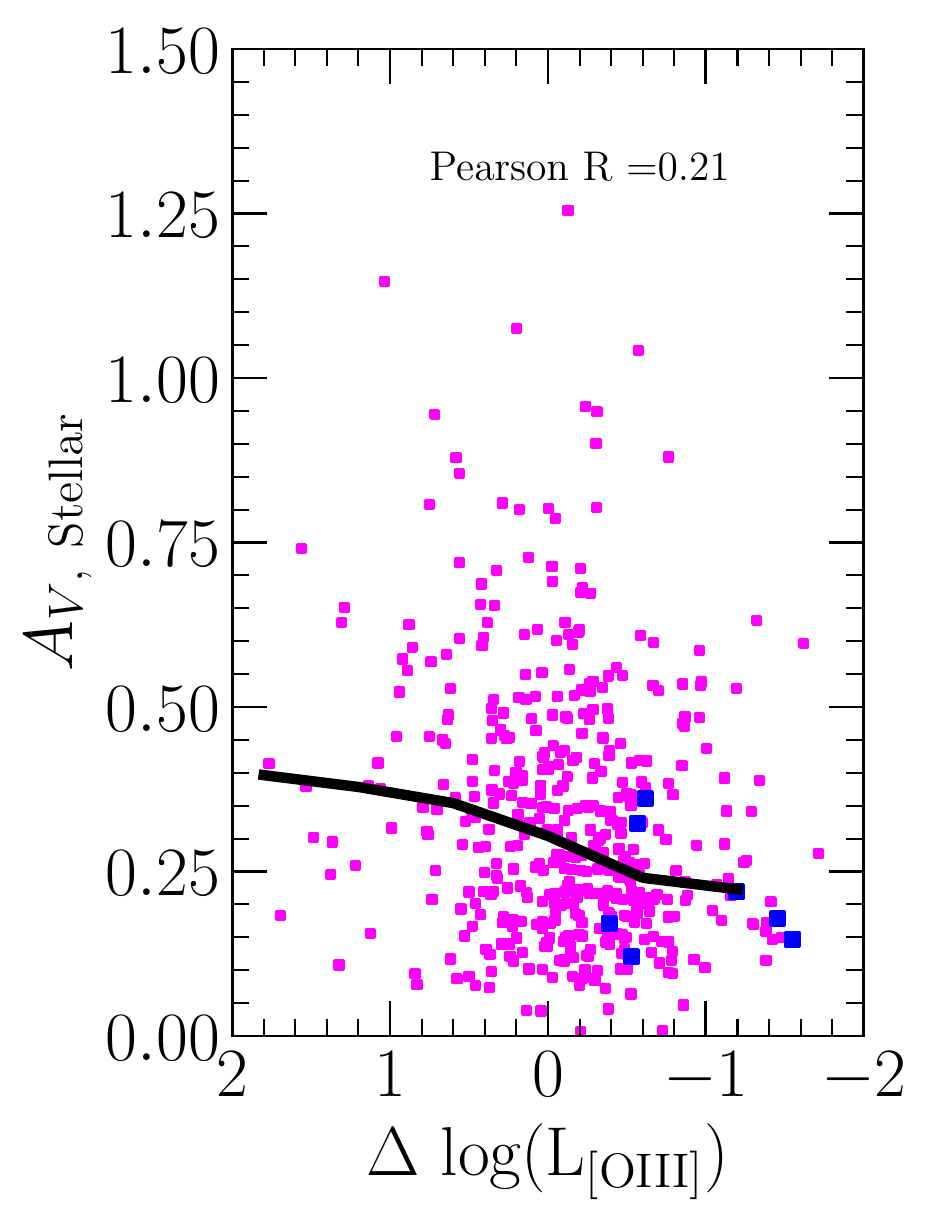}%
            \hfil
            \caption{Potential drivers of the spread of [OIII] luminosity. Horizontal axes go from sources highest above the \lxo\ relation on the left, to the lowest (most underluminous in [OIII]) on the right. Zero on $x$-axis corresponds to the \citet{panessa2006} relation. Left: Host stellar mass as a function of distance from the \lxo\ relation. Middle left: Infrared excess (extra IR emission compared to what is expected from star formation) as a function of distance from the \lxo\ relation. Middle right: Axis ratio (proxy for galaxy inclination and therefore the dust content) as a function of distance from the \lxo\ relation. Right: Stellar continuum attenuation as a function of distance from the \lxo\ relation.  Only unresolved (point) sources (X-ray AGN) are shown. Blue squares are objects with [OIII] SNR below our nominal threshold of $1$, for which we show 1 $\sigma$ upper limit. Only sources with log(sSFR) $>-11$ are included in the middle left panel as the IR excess measurement for sources with log(sSFR)$<-11$ is complicated by dust heating from old stars. Pearson correlation coefficients are provided in each panel.
            \label{fig:dlo3_2}}
        \end{figure*}

\begin{figure}[t!]
        \begin{center}
            \epsscale{1.2}
            \hspace*{-0.2cm}
            
            \plotone{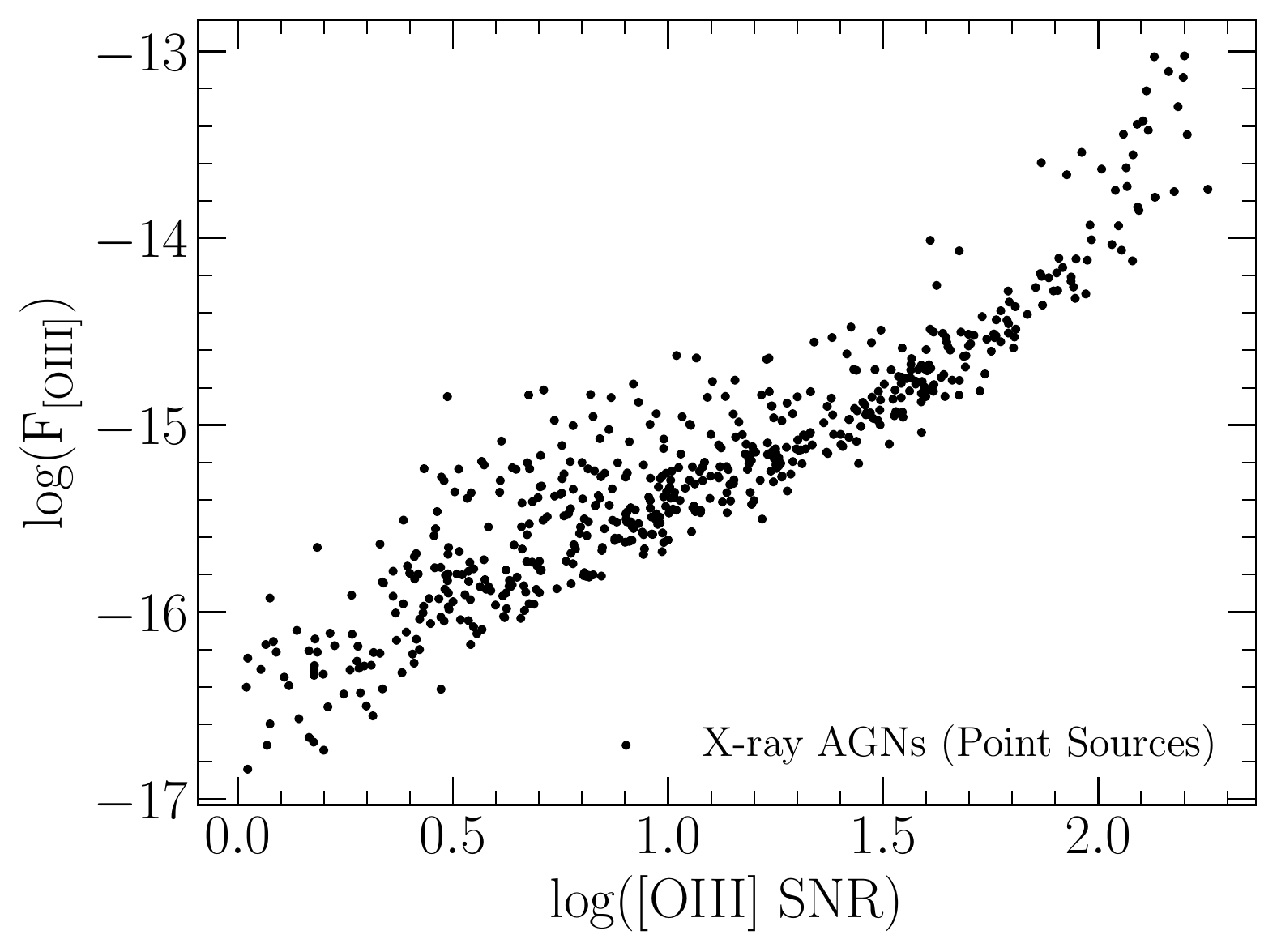}
            
            \caption{[OIII] flux (units of erg s$^{-1}$ cm$^{-2}$) versus the [OIII] SNR. AGNs with low [OIII] SNR, which visually appear as XBONGs, tend to also have weak fluxes, i.e., the low SNR is not the result of a strong continuum contribution (which cannot change the [OIII] flux).
            \label{fig:o3_snr}}

        \end{center}
        \end{figure}

The second phenomenon, continuum swamping of otherwise normal AGN lines (lines of moderate strength), is more relevant in the context of X-ray AGNs with weak or no lines \citep{georgantopoulos2005,caccianiga2007,trump2009,malizia2012}. While it is obvious that the inclusion of host continuum would change the \textit{visual appearance} of the spectrum (the EWs of lines must go down), it is not clear that this would have any effect on the {\it flux} from AGN emission lines. The flux should stay the same---it will just sit on top of a stronger continuum. While strong host contribution may affect how we may classify a spectrum (absorption-line dominated vs.\ emission-line dominated) it should not change the fact that the line flux is still present. Indeed, \citet{moran2002} acknowledge that all of their integrated spectra show emission lines, i.e., none of them actually became an optically dull AGN. This again highlights that one should approach the question of optically dull AGNs from the standpoint of measurements rather than classification. Some studies have pointed out that optically dull AGNs are preferentially found in optically luminous hosts that would more easily overwhelm the AGN lines. Figure \ref{fig:dlo3_2} (left panel) shows that the stellar masses of [OIII]-underluminous AGN are only marginally higher (0.2 dex) than of AGN lying at the \lxo\ relation, and in any case galaxies of high mass have a wide range of  $\Delta L_{\mathrm{[OIII]}}$). Even if the difference had been for some reason larger, we still see no way how the host can do anything but reduce the EW or the SNR of the [OIII] line. We confirm that when the SNR of [OIII] is low it is invariably because the flux is also low---which we show explicitly in Figure \ref{fig:o3_snr}---and not because strong continuum renders it low.

Overall, we conclude that there is very little doubt that [OIII]-underluminous AGN are simply intrinsically less luminous in AGN emission lines (cf.\ \citealt{yuan2004}). Similar conclusions have been reached in studies focused on nearby AGNs where nuclear spectra can be easily isolated and nonetheless find AGNs with weak or no lines \citep{maiolino2003,civano2007}.

\subsubsection{Dust obscuration} \label{sec:dis_dust}

One of the original scenarios that was proposed to explain the weakness of emission lines in XBONGs/optically dull AGNs was that the line emission was obscured by the circumnuclear dust \citep{barger2001,comastri2002}. The idea was particularly attractive around the time when XBONGs started to draw attention because of the concurrent realization that a large population of Compton-thick AGN may explain the spectrum of the X-ray background \citep{comastri2004}. Could these AGNs with atypically weak lines be the manifestation of this Compton-thick population? Subsequent studies have shown that AGNs with weak or no lines do not have Compton-thick gas densities (i.e., log $N_{\mathrm H}>24$), and on the contrary are often unobscured in terms of X-rays (log $N_{\mathrm H}<21.5$), \citep{severgnini2003,civano2007,trump2011b}. In that sense, the gas density distribution of XBONGs/optically dull AGNs does not differ appreciably from that of more typical type 2 AGNs \citep{rigby2006}. It is important to point out that the AGN need not be in the X-ray obscured regime to have a potentially huge effect on the line emission. For the latter, the dust obscuring the ionizing source needs to have a large \textit{covering fraction} (so that the NLR always ``sees'' the ionizing engine through dusty sightlines) rather than having high column density along some sightlines. As a matter of fact, if Compton-thick gas was entirely obscuring the ionizing source there would be no UV photons at all to reach the NLR. Instead, the sort of obscuration that would diminish the [OIII] but not the X-rays would have a high covering factor but relatively low gas dust column, as suggested by \citet{civano2007} and \citet{cocchia2007}. Indeed, to attenuate the [OIII] line by a factor of 10 (2.5 mag) requires only log $N_{\mathrm H}\sim 21.7$, which would not affect the X-ray emission. Therefore, the determination of $N_{\mathrm H}$ or related measures (like the hardness ratio) is entirely irrelevant for establishing if the nuclear dust obscuration is affecting [OIII]. Testing nuclear dust obscuration requires other means.   

If the dust is preventing the UV photons from reaching the NLR, then these photons will heat the dust and give rise to IR emission. If dust obscuration is responsible for AGNs with weak lines, then at a fixed X-ray luminosity the sources that are further below the \lxo\ relation should have a larger fraction of their IR emission associated with the AGN. Distinguishing between the IR emission produced by AGNs as opposed to SF can be challenging, and is best done with detailed mid-IR spectroscopy (e.g., \citealt{kirkpatrick2012}), which is not available for our sources. Fortunately, we can utilize another method to determine the IR excess due to AGN dust heating. \citet{salim2016} have noted that the dust luminosities inferred from the UV-optical SED fitting (without any IR information) generally agree with the dust luminosities determined directly from mid-IR photometry from WISE, except for BPT-selected AGN which exhibit an IR excess that correlates with the [OIII] EW. The IR excess is defined as the difference in the observed IR luminosity and one inferred from the UV/optical SED fitting (i.e., just frmo stars). The IR excess can be reliably measured in star-forming hosts. Galaxies with log sSFR$<-11$ often exhibit an unrelated IR excess due to the dust heating from old populations and are not included in the dataset. In Figure \ref{fig:dlo3_2} (middle left panel) we plot IR excess as a function of \dloiii. A moderate trend exists, but in the direction opposite from what is needed to explain [OIII]-underluminous sources as being preferentially obscured (albeit, IR excess is not available for many of the high $\Delta L_{\mathrm{[OIII]}}$ objects as they have low sSFRs.) Rather, it is the sources above the \lxo\ relation, which are often in the Compton-thick regime, where IR excess seems to be higher on average, suggesting that not only is our particular sightline to the central source obscured for such sources, but there may also be more dust isotropically. No previous study has conclusively determined that the dust associated with the central source is responsible for weak AGN lines and we find no evidence in our sample either. 

      \begin{figure*}[t!]
            \epsscale{1.2}
            \includegraphics[height={2.35in}]{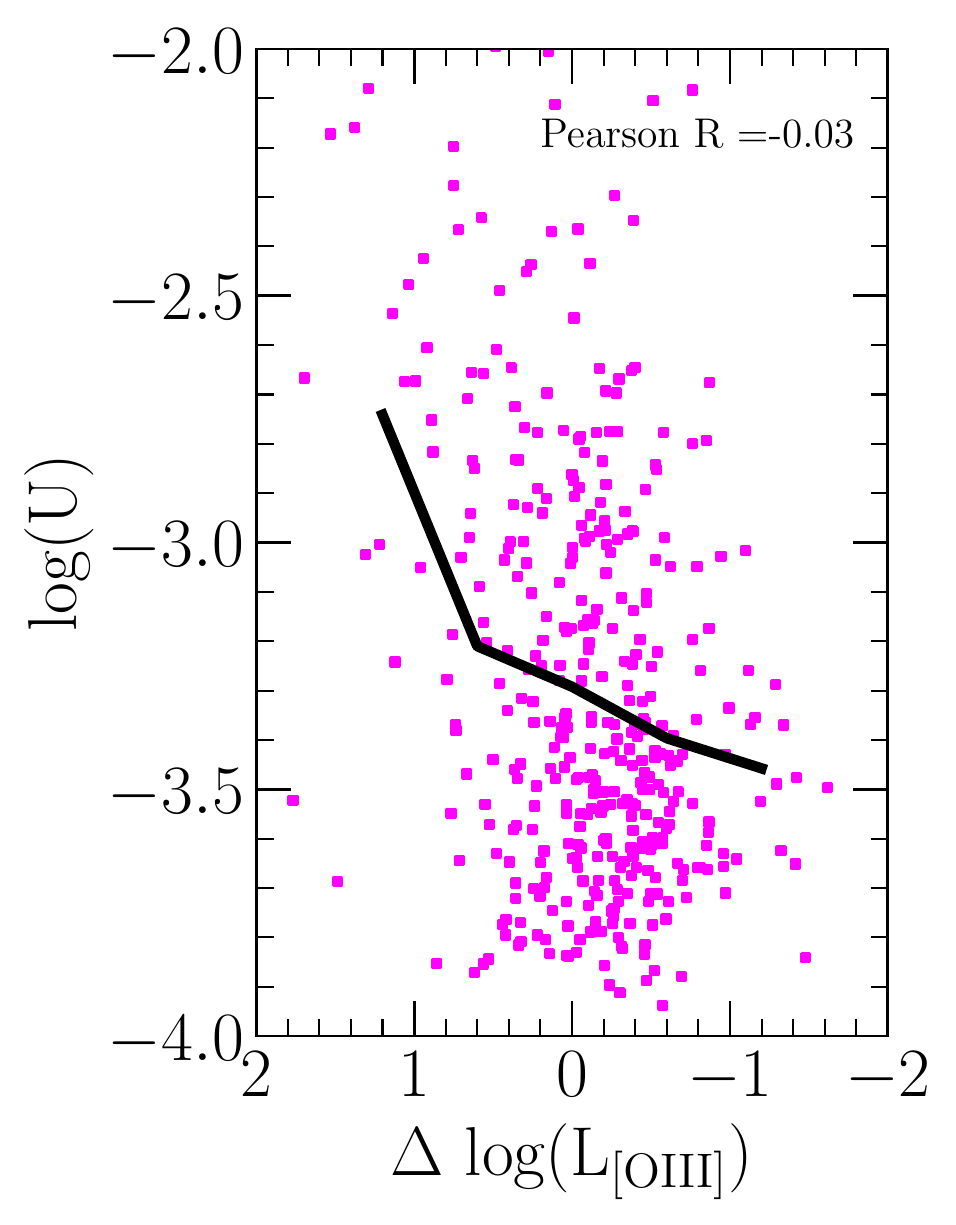}%
            \hfil
            \includegraphics[height={2.35in}]{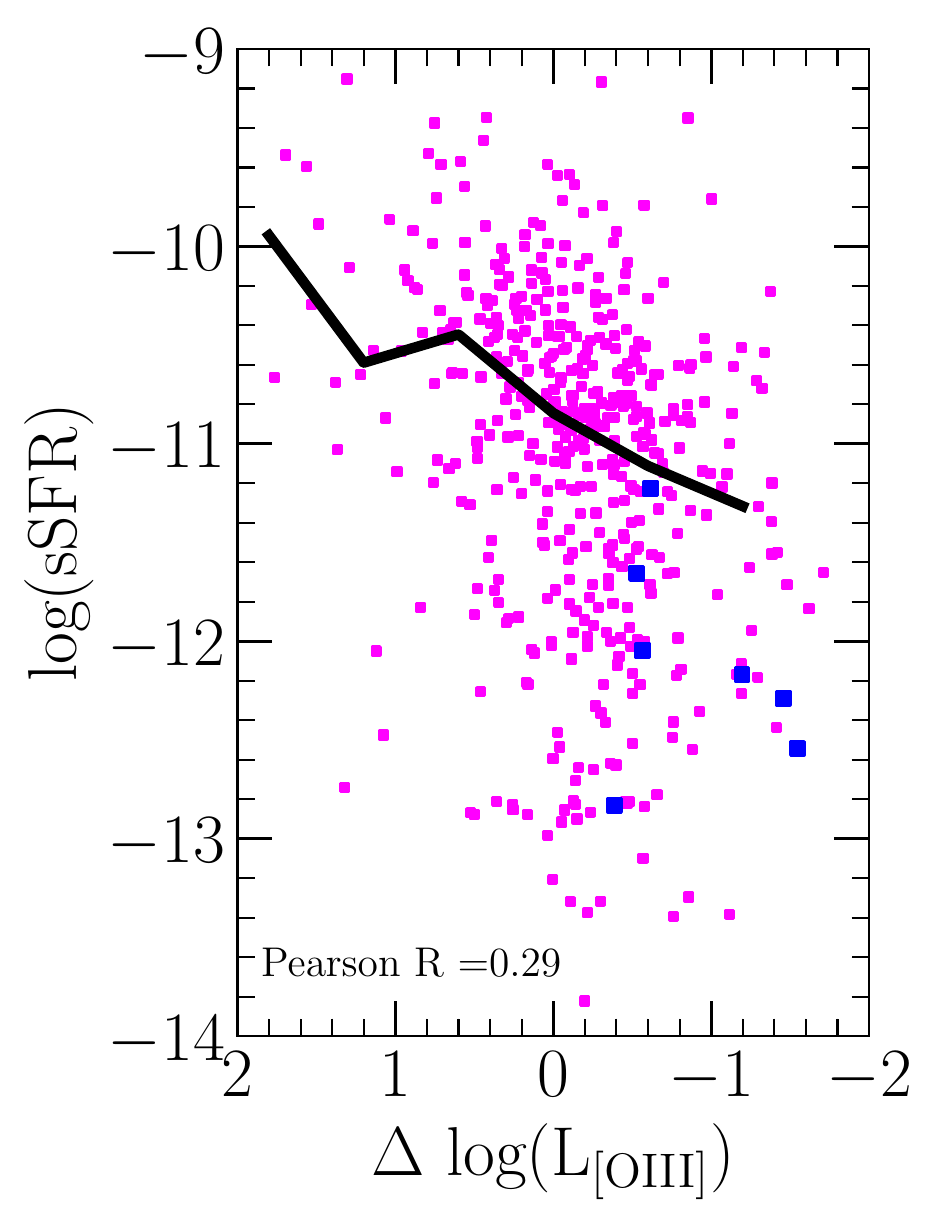}%
            \hfil
            \includegraphics[height={2.3in}]{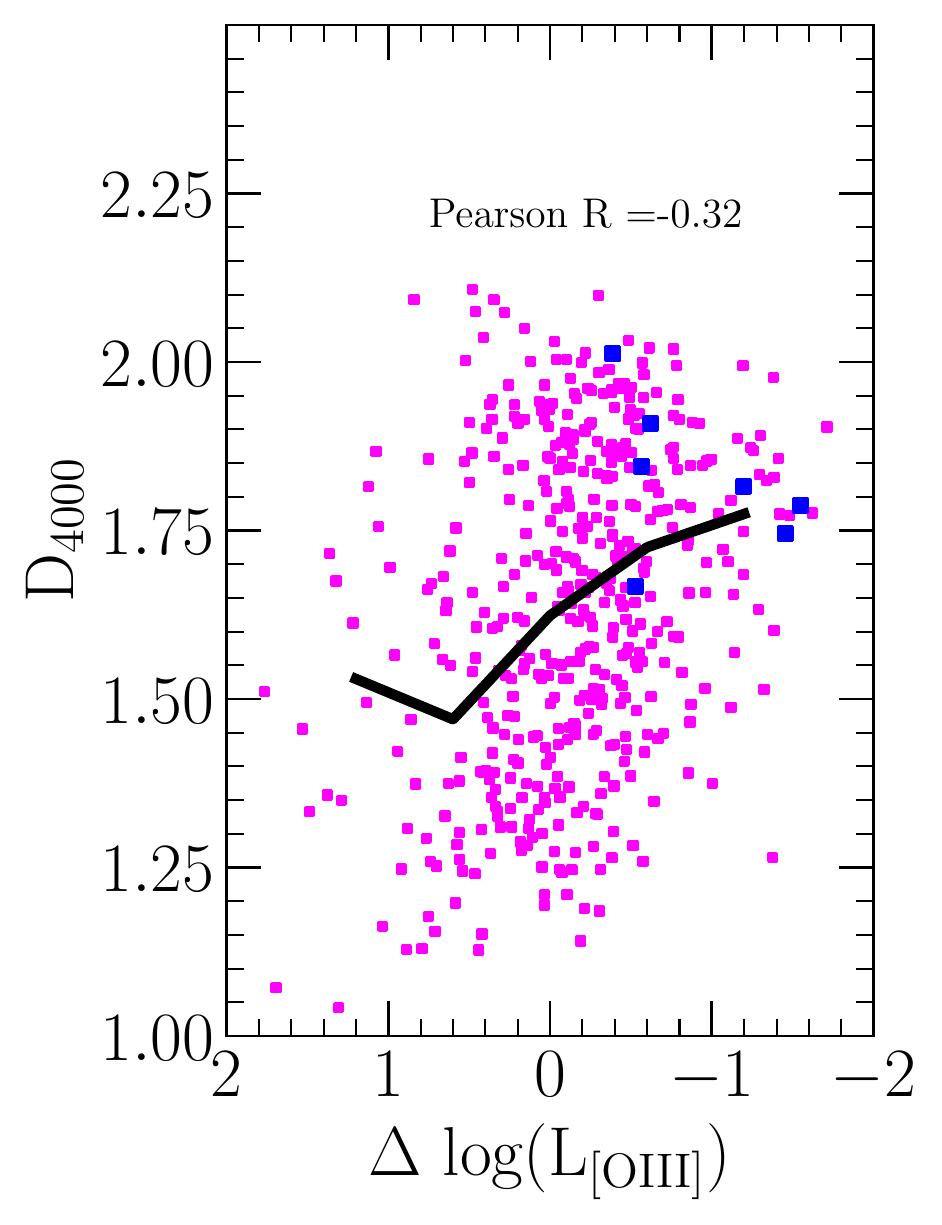}%
            \hfil
            \caption{Potential drivers of the spread of [OIII] luminosity. Horizontal axes go from sources highest above the \lxo\ relation on the left, to the lowest (most underluminous in [OIII]) on the right. Zero on $x$-axis corresponds to the \citet{panessa2006} relation. Left: Ionization parameter as a function of distance from the \lxo\ relation. Middle:  Specific SFR as a function of distance from the \lxo\ relation. Right: D$_{4000}$ as a function of distance from the \lxo\ relation. Only unresolved (point) sources (X-ray AGN) are shown.  Blue squares are objects with [OIII] SNR below our nominal threshold of $1$, for which we show 1 $\sigma$ upper limit. Pearson correlation coefficients are provided in each panel and are the strongest for parameters relating to sSFR and D$_{4000}$, and therefore the amount of gas \label{fig:dlo3_1}}

        \end{figure*}

\citet{rigby2006} examined a different dust scenario, one in which the obscuration of lines is produced by the dust throughout the host galaxy. They support this scenario by noting that in their moderately high redshift sample optically dull AGNs have a wide range of inclinations (including edge-on galaxies that have higher dust columns) whereas ``normal'' type 2 AGNs are more nearly circular. Given that any obscuration of [OIII] by the host would be reflected in the Balmer decrement, and the fact that we already correct the [OIII] emission for such dust attenuation, the very action of correcting [OIII] should remove any underluminous sources.

As this is not the case, the extranuclear dust scenario seems unlikely. However, what if our dust corrections are inadequate or insufficient for some reason? In that case, the sources that appear as [OIII]-underluminous despite the dust correction may be associated with higher dust extinction. To test if that is the case, we follow \citet{rigby2006} in examining the axis ratios, but using much the larger sample available to us. In Figure \ref{fig:dlo3_2} (middle right panel) we show axis ratios as a function of \dloiii. Axis ratio of one indicates a spherical or face-on galaxy. [OIII]-underluminous AGNs are actually slightly rounder than typical ones, suggesting no extra dust content. Because axis ratio is only a proxy for dust attenuation, in Figure \ref{fig:dlo3_2} (right panel) we plot $A_V$ directly as a function of \dloiii. Again, AGNs with weak lines are actually on average somewhat less dusty than the rest.

In summary, neither the nuclear nor the extranuclear (host) dust scenarios for the obscuration of [OIII] emission are supported by our analysis.

\subsubsection{Properties of the ionizing source and the narrow-line region}

In this section we explore two scenarios that would make AGN emission lines be intrinsically weak for reasons unrelated to dust obscuration: (a) RIAF mode of accretion and (b) small NLR covering factors, i.e., the lack of gas to be ionized. 

In a RIAF, the optically thick accretion disk is truncated inward, which reduces the production of UV/optical photons that ionize both the broad and the narrow-line regions \citep{yuan2004}. Thus, RIAFs can naturally explain AGNs with weak lines. Testing if an RIAF is actually responsible for low emission line luminosities is not straightforward. Direct evidence may involve difficult spectrophotometric observations \citep{trump2011b}. Low Eddington ratios are associated with RIAFs in models, and \citet{trump2011a} have shown that their moderately-high redshift sample of ``lineless'' AGN have low Eddington ratios, though there is an overlap in Eddington ratios with regular AGN. Radio emission may be indicative of an RIAF \citep{trump2009}, but the best studied nearby XBONG lacks it \citep{comastri2002}. \citet{rigby2006}, while not explicitly referring to the RIAF scenario, infer from IR observations that the ionizing UV continua of their optically dull AGN appears to be normal, which would suggest the typical mode of accretion (e.g., \citealt{shakurasunyaev1973}). In Figure \ref{fig:dlo3_1} (left panel) we plot the ionization parameter (which represents the ratio of the ionizing photon flux to the gas density, \citealt{netzer1990}) as a function of the distance from the \lxo\ relation. There appears to be a modest trend that  [OIII]-underluminous AGN have lower ionization parameter. However, there is not an indication of the presence of a distinct accretion mode solely among the [OIII]-underluminous AGNs, e.g., a sharp break in ionization parameter. Another, albeit rather qualitative argument why the RIAF scenario may be less favored with our results is the fact that we do not find evidence of there being two population of AGN in regards to emission line properties. Therefore, a unimodal mechanism that exhibits a range in some property appears more natural. 

One such unimodal possibility, which has received very little attention in the literature, is the idea put forward by \citet{trouille2010}. In that scenario, the wide range of \lxorat ratios is the consequence of a wide range of covering factors of the NLR. This is different from a scenario in which dust covers the ionizing region to different extents. If the NLR does not cover the ionizing source sufficiently, it is to be expected that the line emission will be diminished. To our knowledge, no work has tested the NLR covering scenario as an explanation for optically dull AGNs, or generally to understand the diversity of line emission in type 2 AGN. \citet{baskin2005} is one of few studies that explicitly considers the NLR covering factor, along with electron density and ionization parameter, as a factor controlling the [OIII] EW. They demonstrate that the NLR covering factor is the principal factor in driving the [OIII] EW of low-redshift quasars. It is unclear if those results have relevance for type 2 AGN, given that they do not exhibit an AGN continuum like quasars, and therefore the EWs are systematically different. Future studies exploring the role of the NLR covering factor may therefore want to focus on [OIII] flux or luminosity, rather than the EW.

The simplicity of the NLR covering scenario, and the lack of a clear evidence for others, makes it very appealing. But what would be the root cause for NLR covering factors to differ from one AGN to another? \citet{trouille2010} mention the ``complexity in the structure of the NLR'', including how NLR and HII regions are distributed. These scenarios have not been yet tested. Another---perhaps the most straightforward---possibility is that the covering factors differ simply because the gas content in the central regions differs from one galaxy to another. Gas-poor galaxies would have fewer gas clouds, resulting in effectively lower NLR covering factors. Our data do not allow us to examine the gas content directly, so we rely on the specific SFR, which is known to correlate well with the molecular gas fraction \citep{saintonge2017}. In Figure \ref{fig:dlo3_1} (middle panel) we plot specific SFR as a function of \dloiii. We uncover a relatively strong trend (Pearson $R=0.29$) over the entire \dloiii range. We  find a trend of similar strength (Pearson $R=-0.32$) using the fiber D$_{4000}$ (Figure \ref{fig:dlo3_1}, right panel), which unlike the sSFR corresponds only to the region within the spectroscopic fiber. These trends are stronger than for any other parameter we explored. AGNs above the relation (the ones most subject to Compton thick absorption) have sSFRs characteristic of gas-rich main sequence galaxies. Those on the \lxo\ relation have average sSFRs typical of the boundary between the main sequence and the green valley (see Figure \ref{fig:ext_noext} middle panel). The AGNs with no constraints on [OIII] (SNR$<1$) are in the green valley or essentially quiescent (log sSFR$<-12$).

Despite a substantial degree of dispersion in the correlation between sSFR (or D$_{4000}$) and \dloiii, it is evident that AGN with gas-poor hosts tend to lie below the \lxo\ relation. Such AGN obviously need to have the gas available on scales very close to the SMBH to sustain accretion, but they can still be generally gas poor, as is the case for quasars in elliptical galaxies \citep{bahcall1997}. This picture is supported by \citet{lauer2005}, who discuss the strong positive correlation between dust on scales of 100 pc and Seyfert or LINER emission in classical nearby elliptical galaxies. The central gas is what likely fuels an AGN. However, given that in ellipticals the gas (and therefore the dust) is largely absent on NLR scales, a large X-ray luminosity need not be accompanied by what would be the typical [OIII] luminosity that characterizes AGN that have abundant gas. The scales probed by \citet{lauer2005} are much smaller than the SDSS fiber scales in our sample (few kpc), which in turn are comparable in size to NLR (Figure \ref{fig:nlr_sizes}).


A potential objection to the connection between the reduced gas content and low relative [OIII] is that local feedback effects may preferentially remove gas from the NLR for strong AGNs \citep{oosterloo2017, izumi2020, garcia2021, saito2022}. For example, \citet{ellison2021} found that strong (high [OIII] luminosities) AGNs tend to have lower central molecular gas fractions. It is not clear how such results are to be reconciled with what we observe here, but we note that there is substantial uncertainty with respect to how feedback affects the surrounding gas and the physical and the temporal scales associated with those processes.


We conclude that the study of the connection between the gas supply in the NLR and the emission line strengths, especially in a spatially resolved way, may present a way forward not just as the most promising scenario for AGN with weak lines, but for the understanding of the line emission mechanism in general. 


\subsection{Connection between AGNs with weak or no emission lines and AGNs dominated by HII lines}

X-ray AGNs whose spectra show HII-like line ratios are sometimes referred to as ``elusive AGNs'', either together with XBONGs \citep{maiolino2003,caccianiga2007} or in a category of their own \citep{smith2014}. The name ``misclassified AGNs'' has also been applied to them \citep{pons2014} because they appear in the star-forming region of the BPT diagram rather than the expected AGN region. Some of the early studies have already made a connection between AGNs with weak or no emission lines and AGNs dominated by HII lines, the idea being that AGN lines are intrinsically weak but if central SF is present, only the HII lines will be visible in the spectra \citep{moran1996,barger2001,maiolino2003}. \citet{agostino2019} noted that these explanations are conceptually different from the \citet{moran2002} star-formation dilution picture, wherein an AGN with normal lines gets overwhelmed by the HII lines to the extent that the resulting line ratios are like those of SF galaxies. 

Using a sample similar to the one in the current work, \citet{agostino2019} presented evidence that the intrinsic weakness of AGN lines explains these objects better than the SF dilution and therefore calling them ``misclassified'' is somewhat misleading because many of these weak-line AGNs probably would not have had sufficient SNR to be placed on the BPT diagram in the first place. Therefore, AGNs with weak or no emission lines (XBONGs/optically dull AGN) and AGNs dominated by HII lines (elusive AGN) are potentially the same type of objects as far as AGN properties as concerned. It is just that in the latter group the observed [OIII] emission comes from SF, so it is not advisable to mix them with the AGN where that is not the case.

\subsection{Implications of the large scatter in \lxo\ relation on [OIII] as indicator of AGN strength}

Because of the large scatter ($\sim 0.6$ dex) in the \lxo\ relation (or \lxorat ratio), the use of [OIII] as an AGN strength indicator is potentially problematic, as noted in particular by \citet{trouille2010}. We have discussed that above the \lxo\ relation the X-ray luminosity may be subject to gas obscuration, whereas [OIII] is unaffected. On the other hand, below the \lxo\ relation it is [OIII] that is diminished, whereas \lx is indicative of AGN strength. From these considerations, it follows that a preferred indicator of AGN strength could be the [OIII] luminosity when \lx is smaller than what is given by Equation \ref{eq:panessa} and \lx when it is greater than what is given by Equation \ref{eq:panessa}. In practice, this means adopting either [OIII] or X-ray luminosity as a fiducial measure and correcting the values from one side of the \lxo\ relation using Equation \ref{eq:panessa}. More work is required to test the validity of this suggestion.

        \begin{figure}[t!]
        \begin{center}
            \epsscale{1.2}
            \hspace*{-0.2cm}\plotone{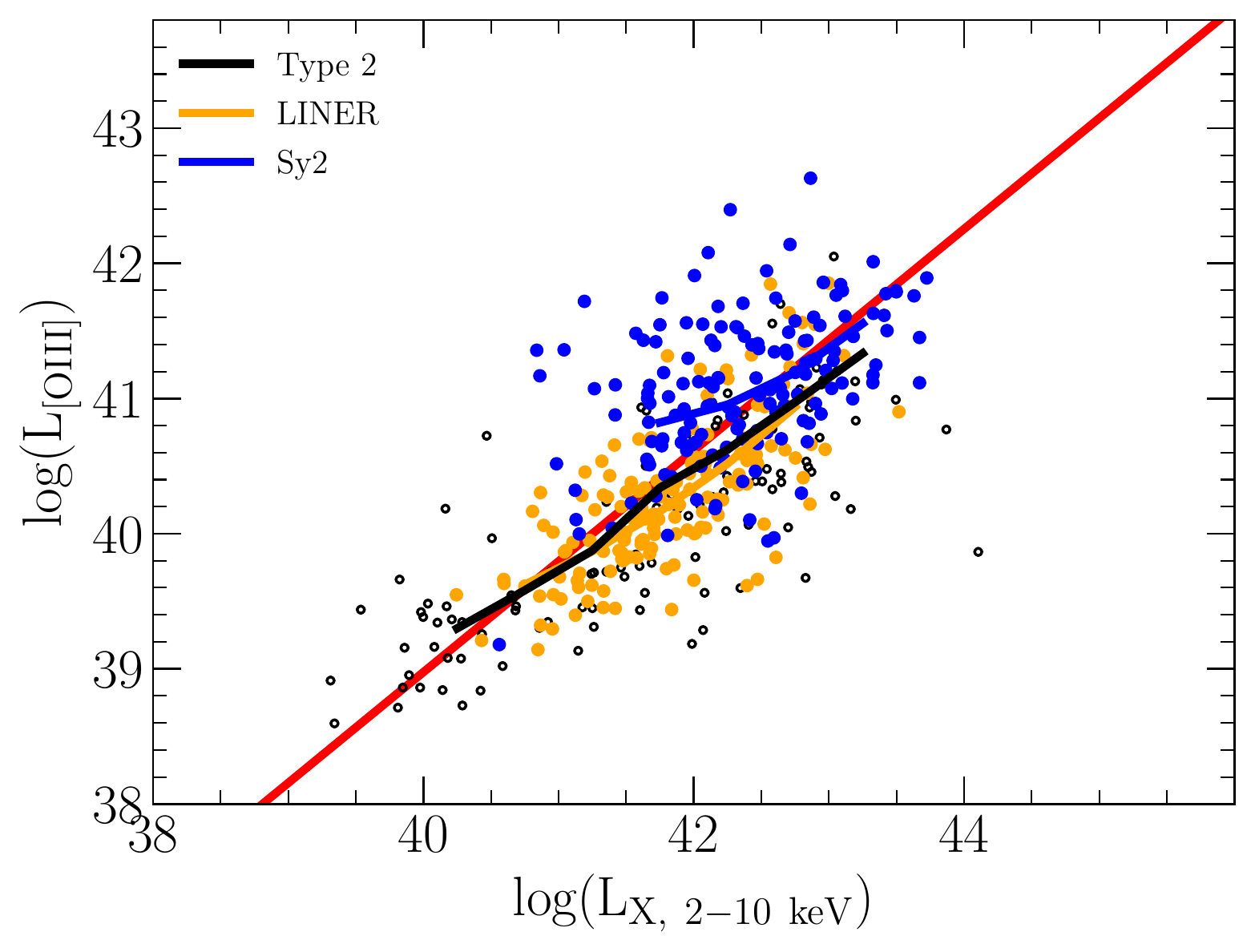}
            \caption{[OIII] luminosity versus X-ray luminosity for Seyfert 2s and LINERs, as classified in \citet{agostino2021}. Only X-ray AGNs are shown here. The median trends are shown as solid lines. Type 2 AGN median trend (black line) is based on Seyfert 2s (blue points), LINERs (orange points) as well as type 2 AGN for which the classification into  Seyfert 2s and LINERs is not available (black open points). LINERS follow the \citet{panessa2006} relation (red line) which is based on well-studied nearby Seyferts.
            \label{fig:lxo3_types}}

        \end{center}
        \end{figure}

\subsection{The AGN nature of emission lines in LINERs?} \label{sec:dis_liners}
    
Since their recognition as a distinct category of emission-line galaxies \citep{heckman1980}, the nature of LINERs and the source of their ionization has been debated extensively in the literature \citep{ho1999,stasinska2008, cidfernandes2011, yan2012, belfiore2016,marquez2017}. Determining whether LINERs are predominantly powered by AGNs or other sources is of critical importance in the census of AGNs and in understanding the potential shortcomings of optical diagnostic diagrams in identifying genuine AGN activity. 
    
Some studies (e.g., \citealt{maiolino2003,caccianiga2007,goulding2009}) include X-ray AGNs with LINER spectra in the ``optically dull'' or ``elusive'' category, such that, like AGNs with weak/no lines or AGNs with HII lines, they require a special explanation. Fundamentally, their view is that a LINER spectrum does not represent a bona fide AGN spectrum like a Seyfert 2 does. \citet{maiolino2003} base that view on the account that LINER emission can in some cases arise in shocks. Subsequent works (e.g., \citealt{stasinska2008, cidfernandes2011, yan2012, belfiore2016}) have argued that LINER-like emission is primarily caused by evolved, post-AGB stars.

In our analysis so far we have focused on all unresolved X-ray sources that are likely to be AGN on account of the excess X-ray emission compared to what is expected from SFR, regardless of their optical AGN classification. We now specifically investigate X-ray AGNs as classified into Seyfert 2s and LINERs (Section \ref{sec:bpt_cl}).  In Figure \ref{fig:lxo3_types} we present the subset of our sample that can be classified as AGN using the BPT diagram and the modified \citet{kauffmann2003} line introduced by \citet{agostino2021}. Furthermore, we color code objects that can be further classified (using the combination of BPT and other lines; Section \ref{sec:bpt_cl}) as either Seyfert 2s (155 objects) or LINERs (152 objects). First, we see that LINERs are quite common among the X-ray selected AGNs, especially if that selection allows the identification of AGNs below the commonly used log \lx $=42$ cut. Next, we see that while Seyferts are rare below log \lx $=41.5$, LINERs and Seyferts overlap over a wide range of X-ray luminosities ($41.5<\log L_{\mathrm X}<43$)---LINERs are clearly not just low-luminosity AGNs. That LINERs can extend to log \lx $=43$ has also been found in \citet{gonzalez-martin2006}. What is remarkable is that the LINERs follow essentially the same \lxo\ relation as the Seyferts (the \lxo\ relation from \citealt{panessa2006} has been constructed mostly from Seyferts). \citet{tanaka2012a} have noted that the presence of a relationship between the X-ray and [OIII] emission of LINERs suggests their AGN nature, as stellar sources (post-AGB stars) would be incapable of producing sufficient X-ray emission. Figure \ref{fig:lxo3_types} confirms this result as it shows a tighter \lxo\ relation for LINERs and that the \lxo\ relation of LINERs is consistent with that of Seyfert 2s, further establishing the case that the optical emission lines in X-ray AGN of LINER type are predominantly powered by an AGN. Interestingly, LINERs, unlike Seyfert 2s, do not tend to scatter to the left and thus are probably less subject to toroidal obscuration compared to Seyfert 2s.

\begin{figure}[t!]
            \epsscale{1.2}
            \hspace*{-0.5cm}\plotone{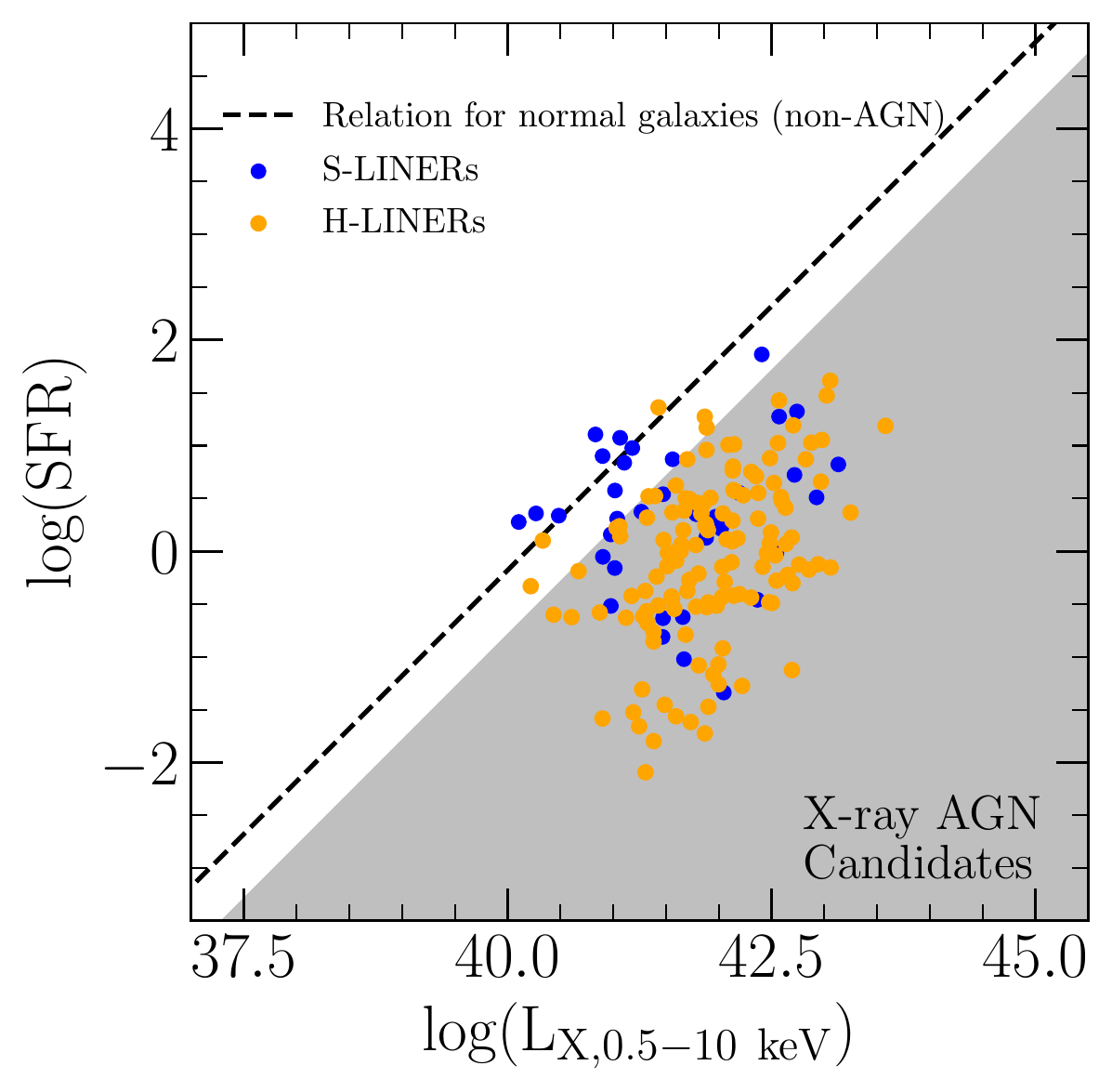}
            \caption{SFRs and full-band X-ray luminosities LINERs split into soft (S-LINER) and hard (H-LINER) subclasses. When detected in X-ray, hard LINERs are generally found to possess an X-ray excess suggestive of an AGN. Soft LINERs, on the other hand, do not always have an X-ray excess. Dashed line and shaded region are the same as in Figure \ref{fig:lxsfr}. \label{fig:lxsfr_liners}}
        \end{figure}

Recently, the analysis of multiple emission lines by \citet{agostino2021} has revealed that galaxies selected as AGN by the BPT diagram consist of Seyfert 2s and two potentially distinct populations of LINERs---soft and hard LINERs. Soft LINERs appear to have softer ionizing spectrum than the more traditional hard LINERs, and lie closer to the \citet{kauffmann2003} demarcation line. Soft and hard LINERs are found in roughly similar numbers in SDSS. If only some LINERs are true AGN, could the distinction be along the soft vs.\ hard line? The answer appears to be partially affirmative. We show the X-ray luminosities and SFRs of hard and soft LINERs in Figure \ref{fig:lxsfr_liners}. We find that hard LINERs, if detected in X-rays, almost always have an X-ray excess with respect to their SFRs (Fig. \ref{fig:lxsfr_liners}), as is the case for Seyferts (Figure \ref{fig:lxsfr}). On the other hand, soft LINERs sometimes do and sometimes do not have an X-ray excess (Fig. \ref{fig:lxsfr_liners}). When they do, they span a similar range in \lx as hard LINERs. Soft LINERs do appear to be some 0.3 dex less luminous in [OIII] than hard LINERs (at fixed \lx), suggesting that whatever process drives the offset from the \lxo\ relation (e.g., the lack of gas supply or the smaller NLR covering factor) is stronger in soft LINERs.

To conclude, both soft and hard LINERs can be true AGNs, but hard LINERs, when detected in X-rays, almost always are.

\citet{cidfernandes2011} have proposed that LINERs powered by AGN and those powered by stellar sources can be distinguished on the basis of H$\alpha$ EW, with the threshold at 3 \AA. We find that 1/3 of our X-ray LINERs fall below this threshold, roughly similar to the overall fraction of LINERs below this threshold. Furthermore, the H$\alpha$ EWs for X-ray LINERs and LINERs altogether are consistent with being drawn from the same distribution (KS-statistic of 0.07 and P-value of 0.32) and that X-ray LINERs are not particularly unique (at least in terms of H$\alpha$ EW) compared to other LINERs. These findings suggest that true (that is, confirmed) AGN LINERs are similarly present at any H$\alpha$ EWs

\section{Conclusions}\label{sec:conc}
In this study, we have used a large sample ($\sim 500$) of low-redshift ($z \sim 0.1$) X-ray AGNs with SDSS spectroscopy to address several open questions surrounding the nature of type 2 AGNs, in particular those that seem to lack the typical optical emission line signatures of AGNs. We have approached the analysis by statistically assessing the properties of the X-ray AGN population in the context of the \lxo\ relation. 

Our major conclusions are the following:

\begin{enumerate}

\item One quarter of the sample of X-ray AGNs looks like pure absorption-line galaxies upon visual inspection of SDSS spectra. However, when line extraction with robust continuum subtraction is utilized (as in the MPA/JHU catalog, \citealt{tremonti2004}), essentially all of these so-called ``optically dull'' AGNs do in fact have measurable emission and therefore it is possible to situate them in the conventional landscape of AGN emission line diagnostics. AGNs thus exhibit various degrees of being underluminous in emission lines rather than belonging to distinct categories (Sections \ref{sec:bpt_res}, \ref{sec:dis_main_res}).

\item AGNs exhibiting weak or apparently no emission lines do not constitute a distinct population of AGNs---there is no dichotomy between ``normal" and so called ``optically dull'' AGNs (i.e., between Seyfert-like and absorption-line spectra; Figures \ref{fig:lxo3_ext_noext} and \ref{fig:deltalo3} and Sections \ref{sec:underlum} and \ref{sec:dis_main_res}).

\item Instead, type 2 AGNs are intrinsically characterized with a very broad (standard deviation of 0.6 dex) distribution of [OIII] luminosities at fixed X-ray luminosity, as noted by \citet{trouille2010}. AGNs visually classified as ``optically dull" ([OIII]-underluminous AGN) are the tail in the unimodal \lxorat distribution (Section \ref{sec:underlum}). 

\item The principal reason why there is large scatter in the \lxo\ relation, and consequently why some AGN appear to be lacking lines, may be related to the gas content of the narrow-line region (cf.\ NLR covering factor of \citealt{trouille2010}). The degree to which an AGN is underluminous in [OIII] correlates with the specific SFR of the host, a proxy for the molecular gas fraction.

\item X-ray AGNs with LINER spectra obey essentially the same \lxo\ relation as Seyfert 2s (have similar [OIII] strength at a given \lx), suggesting that the line emission of X-ray LINERs is produced by AGN activity, rather than a stellar source.

\item Identification of AGNs using emission lines in SDSS is rather complete up to $z \sim 0.15$: it is around 85\% using the BPT diagram alone (with $2 \sigma$ cut on lines), with another 10\% identified based on the high [NII]/Ha ratio as an indication of an AGN (Section \ref{sec:bpt_res}).

\end{enumerate}

Additional findings include:

\begin{enumerate}
\setcounter{enumi}{6}

\item X-ray excess technique allows one to select more extensive samples of genuine X-ray AGNs below the cutoff of log \lx $=42$ (Section \ref{sec:xray_selection}).

\item Extended X-ray sources, which owe bulk of their X-ray emission to hot diffuse gas, can masquerade as a sizable (15-20\%) population of AGNs lacking emission lines (``optically dull" AGN), and must be removed (Section \ref{sec:underlum}). If X-ray extent is not available, certain host properties can be used to identify up to 1/2 of the contaminating sources (Appendix \ref{app:ext}).   

\item ``Host dilution" can be excluded as the reason for AGNs with weak or no emission lines. AGNs with weak lines have low SNR because the line flux/luminosity is low, not because the continuum is strong. In any case, the addition of host continuum would not reduce the measured flux of an AGN line that sits on top of that continuum (Section \ref{sec:dis_aperture}).

\item AGNs with weak [OIII] lines do not have higher AGN IR emission or greater host dust attenuation compared to those on the \lxo\ relation, suggesting that the dust obscuration, either of the ionizing source or by the host galaxy, is not the principal reason for the weak lines. (Section \ref{sec:dis_dust})

\item AGNs with log $L_{\mathrm{X}}> 41.5$ have mostly Seyfert 2 spectra, whereas lower luminosity ones tend to be LINERs. However, both Seyfert 2 and LINERs span large, overlapping ranges in X-ray luminosity, i.e., LINERs are not only low-luminosity AGNs (Section \ref{sec:dis_liners})

\end{enumerate}

\begin{acknowledgments}
This research made use of Astropy, a community-developed core Python package for Astronomy \citep{astropy2013}. The construction of GSWLC used in this work was funded through NASA awards NNX12AE06G and 80NSSC20K0440.
       
CJA and SS thank Katherine Rhode and Liese van Zee for thoughtful comments on the material.

\end{acknowledgments}

\bibliography{refs}

        \begin{figure*}[t!]
            \epsscale{0.8}
             \includegraphics[width={2.05in}]{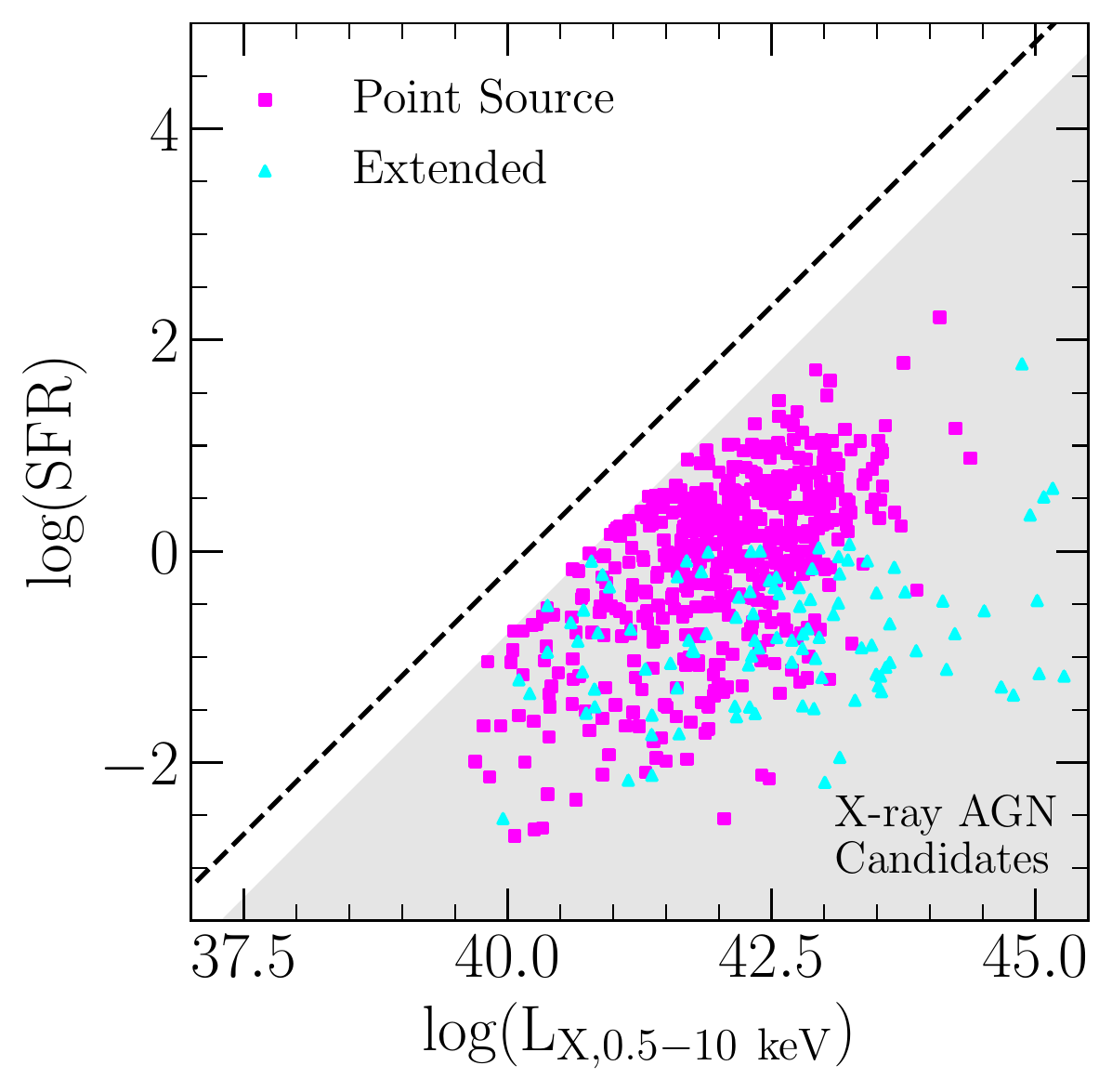}%
            \hfil 
            \includegraphics[width={2.5in}]{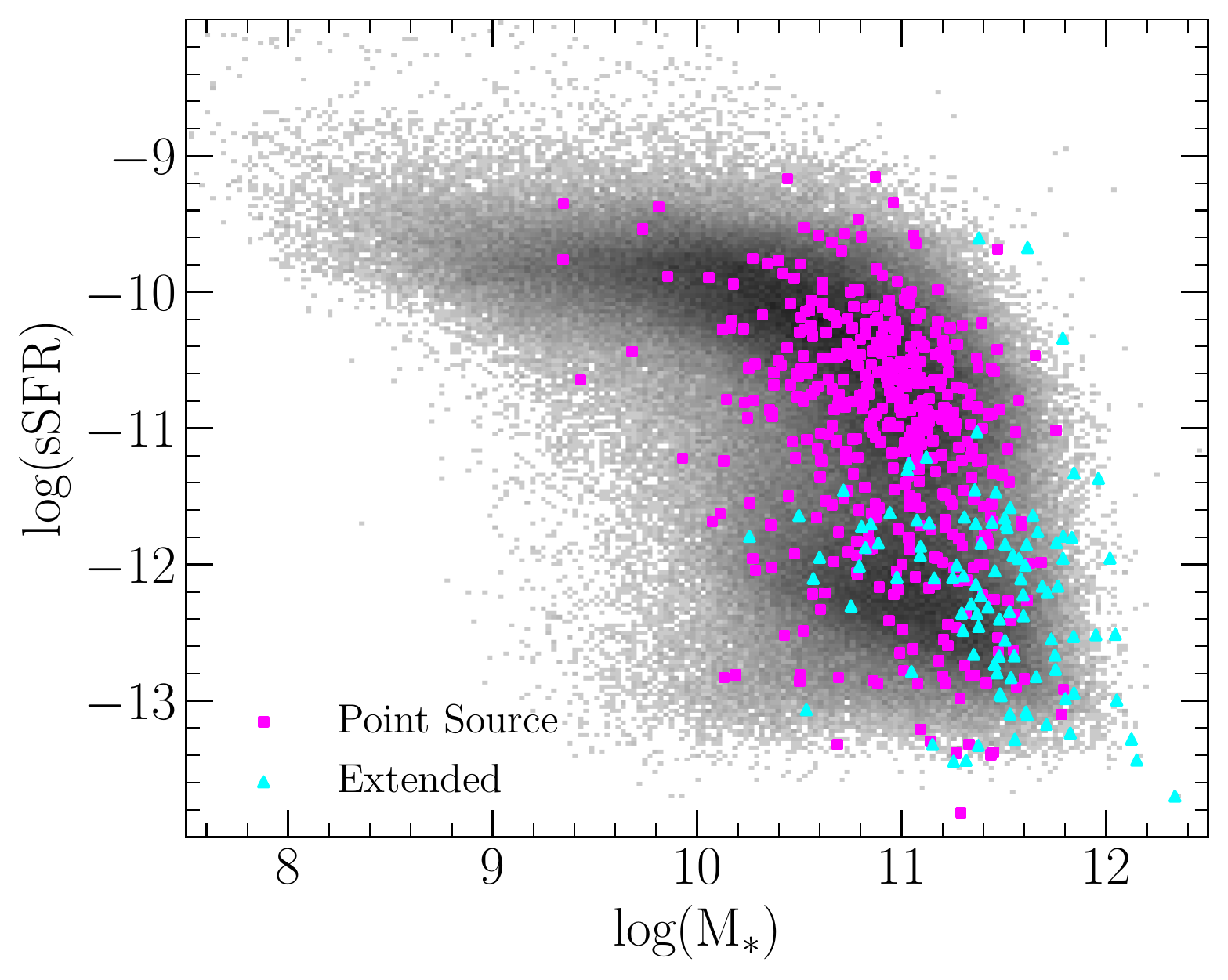}%
            \hfil
            \includegraphics[width={2.3in}]{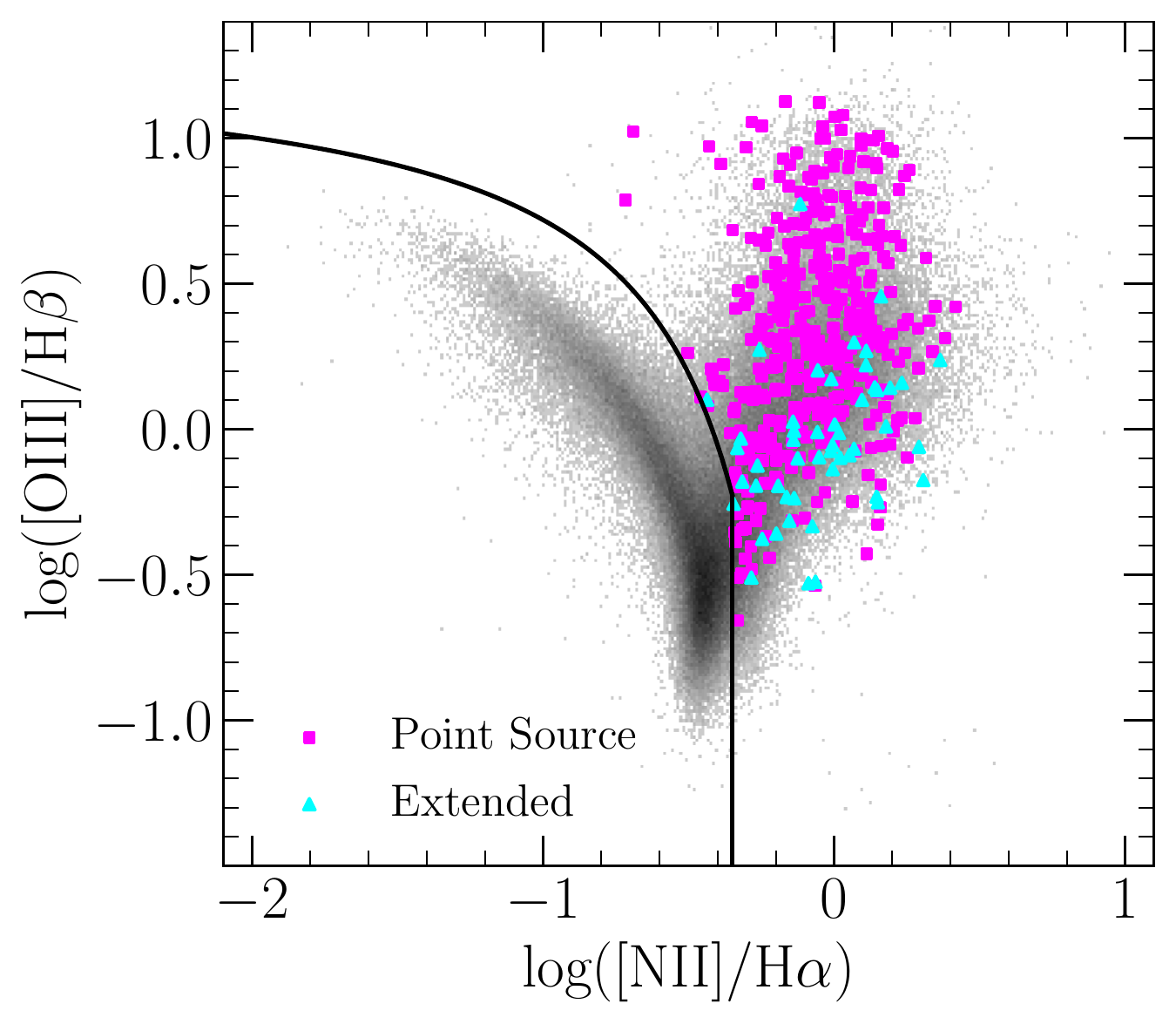}%
            \hfil
            \caption{Properties of X-ray AGN candidates ($z<0.3$) with the extended (resolved) and point-source (unresolved) X-ray emission. The X-ray extent is determined from XMM-Newton observations. Left: X-ray luminosities and SFRs of point sources (magenta squares) and extended sources (cyan triangles). Dashed line and shaded region are the same as in Figure \ref{fig:lxsfr}. Middle: Specific SFR versus stellar mass. Background is made up of all GSWLC-M2 galaxies. Right: BPT diagram of point sources and extended sources. Underlying population is again GSWLC-M2
            \label{fig:ext_noext}}

        \end{figure*}

\appendix

\section{Identification of X-ray sources associated with hot gas emission} \label{app:ext}

As shown in Section \ref{sec:underlum}, the excess in [OIII]-underluminous X-ray AGN candidates was due to an initial inclusion of extended sources, which likely owe the bulk of their X-ray emission to hot diffuse gas, as is often present in galaxy clusters. This provides a cautionary tale to not automatically assume AGN activity when a high X-ray luminosity is measured (see also \citealt{civano2007}). The difficulty of distinguishing cluster emission from genuine AGN emission has been noted by \citet{green2017}, albeit in the context of a search for AGN activity in the centers of clusters. 

We first try to ascertain if environment can be used to identify X-ray sources associated with hot gas emission. We have checked the environment measures provided by \citet{baldry2006} and \citet{blanton2009} and found most of our extended objects to be among rich environments. However, the majority of X-ray sources in such environments are not extended, so these quantitative measures of environment would end up eliminating genuine AGN in addition to extended sources. We also investigated the cluster membership using the catalog of \citet{sdss_clusters} and found that $\sim$60\% of resolved sources in our original sample ($z<0.3$) can be associated with clusters. However, $\sim$30\% of unresolved sources are also associated with clusters, so this environmental information is also by no means sufficient for predicting whether a source will be resolved or unresolved. 

Alternatively, there may exist some host properties associated with extended emission. To evaluate this possibility, we investigated how the extended and point sources compare in terms of their host star formation, stellar mass, X-ray luminosity, and optical emission line properties. In the left panel of Figure \ref{fig:ext_noext} we show SFR versus X-ray luminosity for the extended and the point sources. Extended sources mostly lie below
log(SFR)$< 0$, but there they overlap with point sources unless $\log L_{\mathrm X}>43$. In the middle panel of Figure \ref{fig:ext_noext} we show the specific star formation rates against the stellar mass of the extended sources and the point sources. Extended sources are primarily found in galaxies with low sSFRs (log(sSFR)$<-11$), but so are many point sources, unless $\log M_{*}>11.7$. 

Finally, we investigated the positions of the extended and point sources on the BPT diagram (right panel of Figure \ref{fig:ext_noext}). Most extended sources ($60\%$) cannot be classified with a BPT diagram or even using the [NII]/H$\alpha$ ratio. Those that can be placed on the BPT diagram are mostly found among LINERs, but there they are mixed in with point sources. The fact that some of the extended X-ray sources are AGN according to the BPT diagram emphasizes the fact that some of the extended sources may contain AGNs, in particular radio AGNs. However, their X-ray luminosity is most likely dominated by a non-AGN component.

To summarize, in some cases the host properties can be used to identify extended sources. Specifically, selecting X-ray sources with ($\log L_{\mathrm X}>43$ {\tt AND} log SFR$<0$) {\tt OR} $\log M_{*}>11.7$ identifies 60 out 127 extended sources, while also including 12 out of 473 point sources. Overall yield is therefore not very high ($\sim$50\%), and more work is needed to identify other potential indicators of extended emission that would have a higher yield while not removing a high fraction of point sources.

\end{document}